\documentclass[aps,prd,twocolumn,showpacs,showkeys,amsmath,amssymb]{revtex4-2}
\usepackage{epsfig} 
\usepackage{makecell}
\usepackage{array}
\usepackage{subfigure}
\usepackage{bm}
\usepackage{longtable}
\usepackage{tabularx}
\usepackage{booktabs}
\usepackage{appendix}
\usepackage{caption2}
\usepackage{amsfonts}
\usepackage{graphicx}
\graphicspath{{figs/}}
\usepackage{dcolumn}
\usepackage{siunitx}
\usepackage{bm}
\usepackage{float}
\usepackage{txfonts}
\usepackage{overpic}
\usepackage{feynmf}   
\usepackage{epstopdf}   
\usepackage{slashed}  
\usepackage{multirow}
\usepackage[breaklinks, colorlinks, pdfborder={0 0 0}]{hyperref}

\usepackage[normalem]{ulem}
\usepackage{color}

\allowdisplaybreaks[4]

\begin{document}
	\title{A bootstrap-like improved moment QCD sum rule}%

\author{Jin-Peng Zhang$^1$}	
\email{zhangjp58@mail2.sysu.edu.cn}
\author{Xu-Liang Chen$^1$}
\email{chenxliang@mail2.sysu.edu.cn}
\author{Wei Chen$^{1,\, 2}$}
\email{chenwei29@mail.sysu.edu.cn (Corresponding author)}

\affiliation{$^1$School of Physics, Sun Yat-sen University, Guangzhou 510275, China\\ 
	$^2$Southern Center for Nuclear-Science Theory (SCNT), Institute of Modern Physics, 
	Chinese Academy of Sciences, Huizhou 516000, Guangdong Province, China}
\thanks{Jin-Peng Zhang and Xu-Liang Chen equally contribute to this work.}
\begin{abstract}
QCD sum rules are among the most important non-perturbative tools in hadron physics, with the Laplace sum rule and moment sum rule being the two most commonly used formulations. Despite their widespread application, both approaches have significant shortcomings: the Laplace sum rule relies on subjective criteria to constrain the parameter space, while the conventional moment sum rule cannot extract the coupling constant of the interpolating current and overestimates the ground-state mass. In this work, we propose a bootstrap-like improved moment sum rule framework that resolves these issues simultaneously. Our method explicitly incorporates quark-hadron duality, which introduces the approximation condition on the OPE side and provides a natural a posteriori constraint on the parameters.
We impose dependence conditions on the key parameters as self-consistency conditions of the ``narrow resonance'' ansatz.
As a result, our framework uniquely determines both the optimal value of the duality parameter and the ground-state mass, without invoking any ad hoc or subjective criteria. When we apply the improved moment sum rule to a pseudoscalar $ud\bar{d}\bar{s}$ tetraquark system, the results are in excellent agreement with our previous Laplace sum rule analyses, validating the effectiveness of the proposed scheme. Finally, we apply the method to investigate the $\Sigma\bar{\Sigma}$ molecular system with $I(J^{PC})=0(0^{-+})$. 
The extracted mass is about $2.74\pm 0.12 \ \mathrm{GeV}$, which is much higher than the mass of the $X(2370)$ state. Therefore, our results do not support the interpretation of $X(2370)$ as a $\Sigma \bar{\Sigma}$ molecular state. 
\end{abstract}

\hypersetup{
	citecolor=blue,   
	linkcolor=blue,   
	urlcolor=blue     
}

\maketitle
\pagenumbering{arabic}

\flushbottom

\section{Introduction}
\label{sec:intro}
QCD sum rules provide a well-established non-perturbative framework for studying hadron properties~\cite{Shifman:1978bx,Shifman:1978by,Reinders:1984sr}. The central idea is straightforward: a correlation function is computed in the deep Euclidean region using the operator product expansion (OPE) and then linked to hadron phenomenology via the dispersion relation and quark-hadron duality. Over the years, this approach has been successfully applied to a wide range of hadrons, including conventional mesons and baryons, as well as exotic hadron states such as multiquarks, hybrids, and glueballs~\cite{Chen:2016qju,Lebed:2016hpi,Esposito:2016noz,Guo:2017jvc,Liu:2019zoy,Brambilla:2019esw,Chen:2022asf,Meng:2022ozq,Liu:2024uxn,Wang:2025sic}.

Several variants of the QCD sum rule method have been developed, each emphasizing different aspects of the framework.
These include the Monte Carlo-based approach~\cite{Leinweber:1995fn}, the variable continuum threshold method~\cite{Carvunis:2024koh}, the inverse problem method~\cite{Li:2020ejs,Xiong:2022uwj,Zheng:2026zve}, the inverse matrix method~\cite{Li:2021gsx,Li:2022qul,Li:2024fko,Zhao:2024drr}, the resonance sum rules~\cite{Ou-Yang:2025efp} and so on. Among these variants, the Laplace (or Borel) sum rule (LSR) and the moment sum rule (MSR) are the most widely used ones~\cite{Shifman:1978bx,Reinders:1984sr,Colangelo:2000dp}. 
The Laplace method employs a Borel transform to suppress contributions from excited states and the continuum, whereas the moment method extracts the ground-state mass from moments of the correlation function. Although both approaches share the same theoretical starting point, they treat the phenomenological side differently, which often leads to numerically different results in practice.


The LSR relies on the existence of a ``Borel window'' where both OPE convergence and pole dominance are under control~\cite{Kojo:2006bh}. While this standard procedure often works well, the criteria for selecting the window are not unique. Being largely determined by human intervention, such criteria inevitably introduce procedural uncertainty and subjective bias into the final results. Furthermore, the pole contribution (PC) often varies significantly within the Borel window even when the extracted mass remains stable, raising questions about its physical meaning. The conventional MSR suffers from a different shortcoming. Although the hadron mass can be extracted, the coupling of the current to the ground state remains undetermined. Moreover, the obtained masses are systematically larger than those from LSR, suggesting that the method does not properly isolate the lowest-lying resonance.

In this work, we trace these issues to the treatment of quark-hadron duality. In the conventional moment approach, the OPE spectral density is integrated all the way to infinity, so the extracted mass inevitably receives contributions from higher states. By contrast, the LSR truncates the integral at a scale of the duality parameter via the quark-hadron duality assumption. This observation motivates us to propose an improved moment sum rule (IMSR) that explicitly incorporates quark-hadron duality. The improved framework retains the simplicity of the original moment method while restoring its connection to the ground state, thereby enabling the extraction of both the hadron mass and its coupling.

The core of the IMSR lies in the introduction of the approximation $\delta'_n(s_0,Q^2_0) \simeq  \delta'_{n+1}(s_0,Q^2_0)$ on the OPE side, which mirrors the phenomenological condition $\delta_n(s_h,Q^2_0)\simeq \delta_{n+1}(s_h,Q^2_0)$. This approximation plays a twofold role: it aligns the OPE and phenomenological sides, and it inherently constrains the parameters $(s_0,Q^2_0,n)$ as an a posteriori consistency check.
A further key point is that the ``narrow resonance'' assumption on the phenomenological side implies model self-consistency conditions on the OPE side. On this basis, we define dependence conditions and introduce a tolerance parameter $\epsilon$ to quantify the concentration of the weighted spectral distribution.

Broadly speaking, this framework is closely analogous to the Bootstrap philosophy~\cite{Eden:1966dnq,Chew:1962mpd}: the approximation and dependence conditions effectively act as self-consistency conditions. However, it is important to distinguish our approach from the rigorous Bootstrap program. The latter derives its constraints fundamentally from first principles, such as symmetry, analyticity, and unitarity.

As a test of its validity, we apply the method to the fully light $ud\bar{d}\bar{s}$ tetraquark state. The results confirm its effectiveness. Then we apply it to the $\Sigma\bar{\Sigma}$ molecular state with $I(J^{PC})=0(0^{-+})$, which has been proposed as a possible interpretation of the $X(2370)$~\cite{Ben:2026afy}, a state also predominantly regarded as a glueball candidate~\cite{BESIII:2010gmv,BESIII:2019wkp,BESIII:2023wfi}.

The paper is organized as follows. In Section~\ref{sec:Laplace and moment sum rules}, we briefly review the LSR and conventional MSR, with emphasis on their  limitations. Section~\ref{sec:improved moment sum rules} presents the framework of the improved moment sum rules, where we introduce the approximation $\delta'_n(s_0,Q^2_0) \simeq \delta'_{n+1}(s_0,Q^2_0)$ and the dependence conditions. In Section~\ref{sec:numerical analysis}, we perform numerical analyses for the $ud\bar{d}\bar{s}$ tetraquark state to test its validity, and apply the method to investigate the $\Sigma\bar{\Sigma}$ molecular system with $I(J^{PC})=0(0^{-+})$. In Section~\ref{sec:discussion}, we perform a comparative analysis of the LSR and IMSR, emphasizing their similarities and distinctions. Finally, a summary and outlook are given in Section~\ref{sec:summary}.

\section{Laplace and moment QCD sum rules}\label{sec:Laplace and moment sum rules}
In this section, we briefly review the two mainstream formulations of QCD sum rules that underpin our analysis: the LSR and the conventional MSR. For each approach, we summarize the basic formalism, identify its advantages, and analyze the inherent limitations that motivate the present study. Building on these discussions, we introduce the IMSR method in the following section.

\subsection{Laplace QCD sum rules}
\indent We start from the two-point correlation function
\begin{equation}
	\label{eq:two-point correlation function}
	\begin{aligned}
		\Pi (q^2) = \mathrm{i} \int \mathrm{d}^d x \ \mathrm{e}^{\mathrm{i}  q \cdot x} \langle 0 |
		T [J (x) J
		^{\dagger} (0)] | 0 \rangle,
	\end{aligned}
\end{equation}
where $J(x)$ is the interpolating current with the same quantum numbers as the physical state of interest. At the quark-gluon level, the correlation function can be
computed in the deep Euclidean region $(Q^2 \equiv -q^2 \rightarrow \infty)$ via the method of OPE
\begin{equation}
	\label{eq:ope}
	\begin{aligned}
		\Pi^{\text{OPE}} (q^2) = \sum_i C_i(q^2,\mu) \langle O_i (\mu) \rangle,
	\end{aligned}
\end{equation}
where $\langle O_i (\mu) \rangle$ denote vacuum condensates, $C_i(q^2,\mu)$ are coefficient functions, and $\mu$ is the renormalization scale. The correlation function receives both perturbative and nonperturbative contributions, which encode the short- and long-distance QCD dynamics, respectively. The predictive power of the QCD sum rule approach can be systematically enhanced by incorporating higher-order perturbative and non-perturbative corrections.

On the phenomenological side, the correlation function in Eq.~\eqref{eq:two-point correlation function} can be described by the dispersion relation
\begin{equation}
	\label{eq:dispersion relation}
	\begin{aligned}
		\Pi^{\text{PH}} (q^2) =  \int_{s_N}^{\infty} \mathrm{d} s \frac{ (q^2)^n\rho^{\text{PH}} (s)}{s^n (s - q^2 -\mathrm{i}0^+)} + \sum^{n - 1}_{i = 0} a_i
		(q^2)^i ,
	\end{aligned}
\end{equation}
where $s_N$ is the normal threshold and $a_i $ are coefficients of  subtraction terms. The hadronic spectral function can be parametrized by using the ``narrow resonance'' assumption
\begin{equation}
	\label{eq:narrow resonance}
	\begin{aligned}
		\rho^{\text{PH}} (s) =  f_X^2 \delta (s - m_X^2) + \rho^h(s)\theta(s-s_h),
	\end{aligned}
\end{equation}
in which $m_X$ and $f_X$ are the hadron mass and coupling of the ground state, respectively, and $\rho^h(s)$ represents contributions from the continuum and excited states with $s_h$ denoting their threshold.

The cornerstone of the QCD sum rule approach is the assumption of quark-hadron duality, which posits that for $s$ above the duality parameter $s_0$, the phenomenological spectral density $\rho^h(s)$ coincides with its OPE counterpart, $\rho^{\text{OPE}}(s)= \frac{1}{\pi}\operatorname{Im}\Pi^{\text{OPE}}(s)$. Local duality can be formulated as
\begin{equation}
	\label{eq:loaca duality}
	\begin{aligned}
		\rho^h(s)=\frac{1}{\pi} \operatorname{Im} \Pi^{\text{OPE}}(s) \theta(s-s_0),
	\end{aligned}
\end{equation}
while global duality takes the form
\begin{equation}
	\label{eq:global duality}
	\begin{aligned}
		\int^{\infty}_{s_h}\mathrm{d}s\frac{\rho^h(s)}{s-q^2}= \int^{\infty}_{s_0}\mathrm{d}s\frac{\rho^{\text{OPE}}(s)}{s-q^2}.
	\end{aligned}
\end{equation}
Both the lack of knowledge of $\rho^h(s)$~\cite{Lucha:2007pz} and ``duality violation''~\cite{Shifman:2000jv,Shifman:2010zza} will inevitably introduce theoretical uncertainties.

Then the Borel transform is applied to both the phenomenological and QCD sides
\begin{equation}
	\label{eq:Borel}
	\begin{aligned}
		\hat{B}=\lim_{\substack{-q^2 , n \to \infty  \\ -q^2/n = M^2_B}} \frac{1}{\Gamma(n+1)}(-q^2)^{n+1}\left(\frac{\mathrm{d}}{\mathrm{d}q^2}\right)^n,
	\end{aligned}
\end{equation}
which will significantly suppress the contributions from higher excited states, improve the convergence of OPE series, and eliminate the unknown subtraction terms in Eq.~\eqref{eq:dispersion relation}, according to the LSR perspective. Subsequently, the LSR is obtained as
\begin{equation}
	\label{eq:mass of laplace}
	\begin{aligned}
		f^2_X \ \mathrm{e}^{-m^2_X/M^2_B} =\int_{s_N}^{s_0} \rho^\text{OPE}(s) \ \mathrm{e}^{-s/M^2_B} \ \mathrm{d}s,
	\end{aligned}
\end{equation}
in which $M^2_B$ is the Borel parameter. Differentiating both sides of Eq.~\eqref{eq:mass of laplace} with respect to $\frac{-1}{M^2_B}$, the hadron mass $m_X$ of the lowest-lying resonance can be extracted to be
\begin{equation}
	\label{eq:mass of laplace 2}
	\begin{aligned}
		m_X(s_0,M^2_B)=\sqrt{\frac{L_1(s_0,M^2_B)}{L_0(s_0,M^2_B)}},
	\end{aligned}
\end{equation}
where
\begin{equation}
	\label{eq:laplace moment}
	\begin{aligned}
		L_k(s_0,M^2_B)=\int^{s_0}_{s_N}\rho^{\text{OPE}}(s) s^k \mathrm{e}^{-s/M^2_B} \ \mathrm{d}s.
	\end{aligned}
\end{equation}
With a fixed normal threshold $s_N$, the hadron mass in Eq.~\eqref{eq:mass of laplace 2} depends only on the duality parameter $s_0$ and the Borel parameter $M^2_B$. These two parameters are conventionally constrained by imposing two criteria: the convergence of the OPE series in Eq.~\eqref{eq:ope} and the dominance of the pole over the continuum in Eq.~\eqref{eq:narrow resonance}. However, there is no unique definition and unified standard for both the above criteria, which usually result in confused subjective choices in practice.
Consequently, the value of $s_0$, and hence the hadron mass $m_X$, is not uniquely determined, introducing a certain degree of ambiguity and potential systematic bias into the final results. 

On the other hand, the definition of pole contribution (PC) itself is questionable  
\begin{equation}
	\begin{aligned}
		\text{PC}(s_0,M^2_B) := \frac{L_0(s_0,M^2_B)}{L_0(\infty,M^2_B)},
	\end{aligned}
\end{equation}
which receives contributions from all terms contained in the correlation function.
A more fundamental issue, however, is the ambiguous physical meaning of the PC itself. Within the Borel window, we observe a stark contrast: while the extracted hadron mass remains stable, the PC varies considerably. This presents an internal inconsistency: if the PC were a physically meaningful quantity, it should not be highly sensitive to the unphysical parameter $M^2_B$. Moreover, even if one assumes the existence of a stable interval for the PC with respect to $M^2_B$, this interval must necessarily overlap with the stability window of the hadron mass. Otherwise, their stability behaviors would be contradictory.

\subsection{The conventional moment QCD sum rules}
The MSR follows essentially the same procedure as the LSR, differing only slightly on the phenomenological side. Instead of performing the Borel transformation, one defines the following moments by taking derivatives of the correlation function $\Pi(q^2)$ in the Euclidean region  $Q^2 \equiv -q^2 >0$~\cite{Reinders:1984sr}
\begin{equation}
	\label{eq:def of moment}
	\begin{aligned}
		M_n(Q_0^2)= & \frac{1}{\Gamma(n+1)}\left( -\frac{\mathrm{d}}{\mathrm{d} Q^2} \right)^n \Pi(Q^2)|_{Q^2=Q^2_0}.
	\end{aligned}
\end{equation}
On the quark-gluon side, the moment $M_n(Q^2_0)$ can be obtained from the OPE, just as in the LSR
\begin{equation}
	\label{eq:OPE def of old moment}
	\Pi^{\text{OPE}}(q^2)
	= \int_{s_N}^{\infty} \frac{\rho^{\text{OPE}}(s)}{s-q^2}\mathrm{d}s,
	\
	M_n^{\text{OPE}}(Q_0^2)
	= \int_{s_N}^{\infty} \frac{\rho^{\text{OPE}}(s)}{(s+Q_0^2)^{n+1}}\mathrm{d}s .
\end{equation}

Using Eq.~\eqref{eq:narrow resonance}, one can write the moment function on the phenomenological side as
\begin{equation}
	\label{eq:easy of old moment}
	\begin{aligned}
		M_n^{\text{PH}}(Q_0^2) & =\frac{f^2_X}{(m^2_X+Q^2_0)^{n+1}}[1+\delta_n(s_h,Q^2_0)], \\
		\delta_n(s_h,Q^2_0) & :=\frac{(m^2_X+Q^2_0)^{n+1}}{f^2_X} \int_{s_h}^{\infty} \frac{\rho^h(s)}{(s+Q^2_0)^{n+1}}\mathrm{d}s,
	\end{aligned}
\end{equation}
in which $\delta_n(s_h,Q^2_0)$ contains the contributions of higher excited states. To eliminate the coupling $f_X$ in Eq.~\eqref{eq:easy of old moment}, we consider the following ratio of the moments
\begin{equation}
	\label{eq:PH side of moment}
	\begin{aligned}
		r^{\text{PH}}(n,Q^2_0)=\frac{M^{\text{PH}}_n(Q_0^2)}{M^{\text{PH}}_{n+1}(Q_0^2)}=(m^2_X+Q^2_0)\frac{1+\delta_n(s_h,Q^2_0)}{1+\delta_{n+1}(s_h,Q^2_0)}.
	\end{aligned}
\end{equation}
On the OPE side, the same ratio can be considered as
\begin{equation}
	\begin{aligned}
		r^{\text{OPE}}(n,Q^2_0)=\frac{M^{\text{OPE}}_n(Q_0^2)}{M^{\text{OPE}}_{n+1}(Q_0^2)}.
	\end{aligned}
\end{equation}
Both sides should yield the same result
\begin{equation}
	\label{eq:equ in moment}
	\begin{aligned}
		r^{\text{OPE}}(n,Q^2_0)=r^{\text{PH}}(n,Q^2_0).
	\end{aligned}
\end{equation}
The moment approach also relies on an additional approximation. According to conventional wisdom, for sufficiently large $n$, one expects that $\delta_n(s_h,Q^2_0) \simeq \delta_{n+1}(s_h,Q^2_0)$ for convergence. Hence, the rightmost part of Eq.~\eqref{eq:PH side of moment} simplifies to
\begin{equation}
	\label{eq:easy PH side of moment}
	\begin{aligned}
		r^{\text{PH}}(n,Q^2_0)= m^2_X+Q^2_0.
	\end{aligned}
\end{equation}
Then the hadron mass of the lowest-lying resonance $m_X$ can be immediately extracted as
\begin{equation}
	\label{eq:mass of old moment}
	\begin{aligned}
		m_X(n,Q^2_0)=\sqrt{r^{\text{OPE}}(n,Q^2_0)-Q_0^2}.
	\end{aligned}
\end{equation}

Both the MSR and LSR can be used to extract the ground state hadron mass. However, Eq.~\eqref{eq:easy of old moment} reveals a drawback of the MSR approach: the lack of knowledge of both $\rho^h(s)$ and $\delta_n(s_h,Q^2_0)$ prevents the determination of the coupling $f_X$. A more striking issue is the systematic discrepancy between the two methods: the conventional MSR consistently yields a larger mass prediction than LSR.

Motivated by these observations, we introduce in the next section an improved version of the moment sum rules that simultaneously resolves all of these issues.

\section{The improved moment QCD sum rules}\label{sec:improved moment sum rules}
In this section, we establish the formalism of the improved moment QCD sum rules. We discuss its key idea and basic procedure, while deferring the illustrative examples to the next section.
\subsection{Motivation and core idea}
To begin with, it is instructive to examine why the conventional MSR systematically yields a higher hadron mass compared with the result of LSR.

Labeling the correlation function as ``OPE'' and ``PH'' in Sec.~\ref{sec:Laplace and moment sum rules}, we have explicitly indicated that the approximation $\delta_n(s_h,Q^2_0) \simeq \delta_{n+1}(s_h,Q^2_0)$ is imposed solely on the phenomenological side, leaving the OPE side intact. A comparison of Eq.~\eqref{eq:mass of laplace} with Eq.~\eqref{eq:OPE def of old moment} reveals a crucial difference: in the MSR, the integration over the OPE spectral density $\rho^{\text{OPE}}(s)$ extends to infinity, whereas in the LSR, the integral is effectively cut off at the duality parameter $s_0$. This distinction implies that the ground state hadron mass $m_X$ extracted from the MSR inevitably receives contamination from higher excited states, which naturally shifts the extracted mass upward. The difference in the upper limit of integration ultimately originates from the implementation of quark-hadron duality, which is explicitly enforced in the LSR but not in the moment approach.

\begin{figure*}[htbp]
	\centering
	\includegraphics[width=0.5\textwidth]{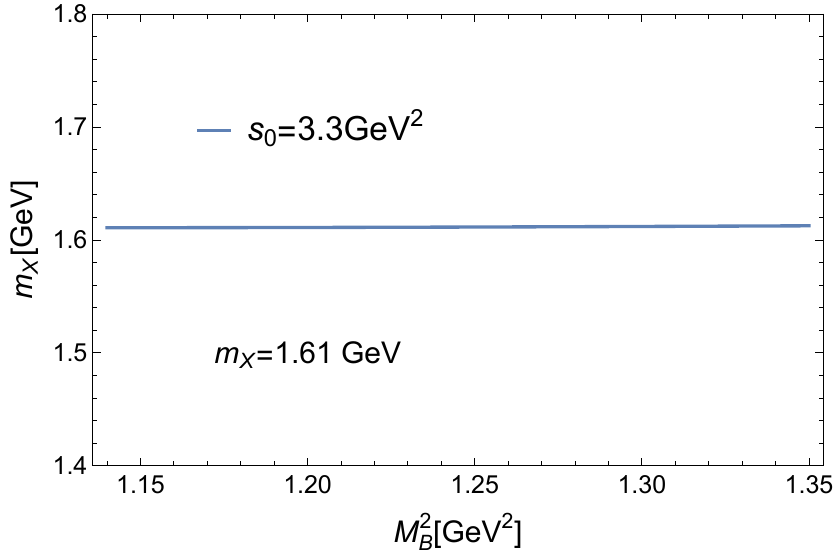}
	\caption{Mass extracted from the LSR with $s_0=3.3\,\mathrm{GeV}^2$.}
	\label{fig:rho mass in Laplace}
\end{figure*}

\begin{figure*}[htbp]
	\centering
	\subfigure[]{
		\includegraphics[width=0.45\textwidth]{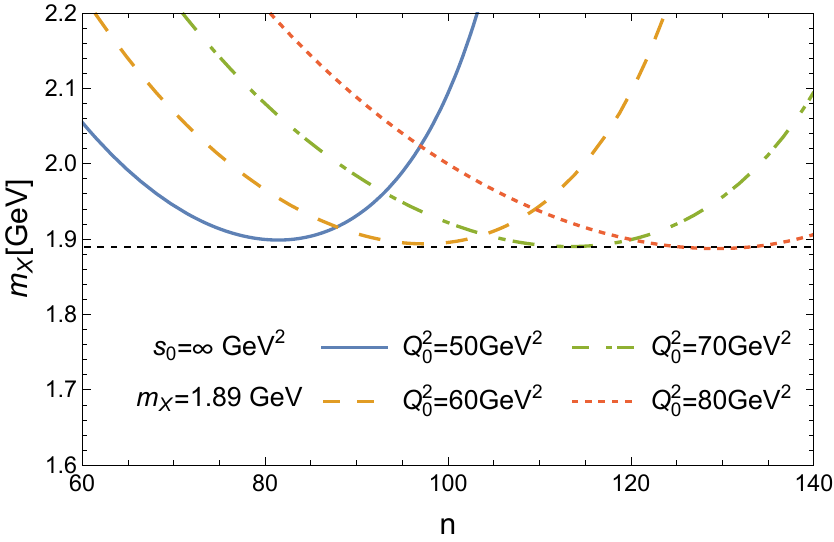}
		\label{fig:rho mass in old moment}
	}
	\subfigure[]{
		\includegraphics[width=0.45\textwidth]{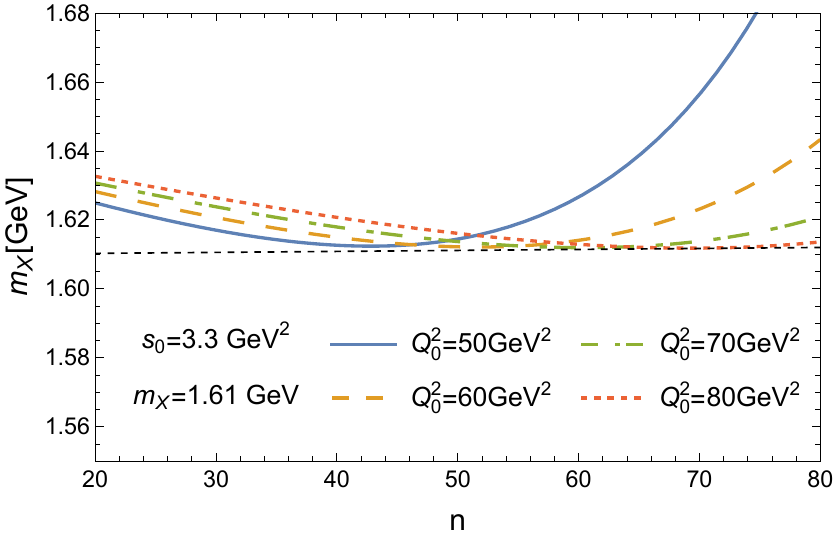}
		\label{fig:rho mass in s0 moment}
	}
	\caption{Masses extracted from the MSR with the upper limit of integration in Eq.~\eqref{eq:OPE def of old moment} as infinity in Fig.~\ref{fig:rho mass in old moment} and $3.3\,\text{GeV}^2$ in Fig.~\ref{fig:rho mass in s0 moment}.}
	\label{fig:rho mass in moment}
\end{figure*}

An important observation is that the mass discrepancy between the LSR and MSR can be resolved simply by adopting the same duality parameter $s_0$. To illustrate this, we compare in Figs.~\ref{fig:rho mass in Laplace} and~\ref{fig:rho mass in moment} the numerical results obtained from the two methods for the pseudoscalar tetraquark state $ud\bar d\bar s$, with its OPE expression given in Eq.~\eqref{eq:OPE of P6}. When each method is analyzed with its own independently optimized parameters, the MSR mass is about 0.3 GeV higher than the LSR mass. However, when both methods are evaluated with the same duality parameter, the extracted masses converge to the same value.

Given the above discussion, it is essential that quark-hadron duality be incorporated into the MSR as well, in order to ensure an accurate extraction of the lowest-lying hadron mass. Accordingly, the moment of the correlation function in Eq.~\eqref{eq:OPE def of old moment} can be written as
\begin{widetext}
	\begin{equation}
		\label{eq:new def of old moment}
		\begin{aligned}
			M_n^{\text{OPE}}(Q_0^2)  = \int_{s_N}^{s_0} & \frac{\rho^{\text{OPE}}(s)}{(s+Q^2_0)^{n+1}} \mathrm{d}s  + \int_{s_0}^{\infty} \frac{\rho^{\text{OPE}}(s)}{(s+Q^2_0)^{n+1}} \mathrm{d}s = \tilde{M}^{\text{OPE}}_n(s_0,Q_0^2)+M'^{\text{OPE}}_n(s_0,Q_0^2)\, ,\\ 
			\tilde{M}^{\text{OPE}}_n(s_0,Q_0^2)& :=  \int_{s_N}^{s_0} \frac{\rho^{\text{OPE}}(s)}{(s+Q^2_0)^{n+1}} \mathrm{d}s,  \quad
			M'^{\text{OPE}}_n(s_0,Q_0^2) :=  \int_{s_0}^{\infty} \frac{\rho^{\text{OPE}}(s)}{(s+Q^2_0)^{n+1}} \mathrm{d}s\, .
		\end{aligned}
	\end{equation}
\end{widetext}
After incorporating quark-hadron duality, the moment function decomposes into two contributions $\tilde{M}^{\text{OPE}}_n(s_0,Q_0^2)$ and $M'^{\text{OPE}}_n(s_0,Q_0^2)$. According to Eq.~\eqref{eq:global duality}, $M'^{\text{OPE}}_n(s_0,Q^2_0)$ satisfies:
\begin{equation}
	\label{eq:new duality}
	\begin{aligned}
		M'^{\text{OPE}}_n(s_0,Q^2_0)
		=\int_{s_h}^{\infty} \frac{\rho^h(s)}{(s+Q^2_0)^{n+1}} \mathrm{d}s.
	\end{aligned}
\end{equation}
Subsequently, we can rewrite Eq.~\eqref{eq:equ in moment} as
\begin{equation}
	\label{eq:equ in new moment}
	\begin{aligned}
		\frac{\tilde{M}^{\text{OPE}}_n(s_0,Q_0^2)+M'^{\text{OPE}}_n(s_0,Q_0^2)}{\tilde{M}^{\text{OPE}}_{n+1}(s_0,Q_0^2)+M'^{\text{OPE}}_{n+1}(s_0,Q_0^2)}\\ =
		(m^2_X+Q^2_0)\frac{1+\delta_n(s_h,Q^2_0)}{1+\delta_{n+1}(s_h,Q^2_0)}.
	\end{aligned}
\end{equation}
To maintain a uniform formalism, we define the function $\delta'_n(s_0,Q^2_0)$ as
\begin{equation}
	\label{eq:define of delta 2}
	\begin{aligned}
		\delta'_n(s_0,Q^2_0) := \frac{M'^{\text{OPE}}_n(s_0,Q^2_0)}{\tilde{M}^{\text{OPE}}_n(s_0,Q^2_0)},
	\end{aligned}
\end{equation}
so that Eq.~\eqref{eq:equ in new moment} can be expressed as
\begin{equation}
	\label{eq:easy equ in new moment}
	\begin{aligned}
		\frac{\tilde{M}^{\text{OPE}}_n(s_0,Q_0^2)}{\tilde{M}^{\text{OPE}}_{n+1}(s_0,Q_0^2)} \frac{1+\delta'_n(s_0,Q_0^2)}{1+\delta'_{n+1}(s_0,Q_0^2)} \\
		=(m^2_X+Q^2_0)\frac{1+\delta_n(s_h,Q^2_0)}{1+\delta_{n+1}(s_h,Q^2_0)}.
	\end{aligned}
\end{equation}

In the conventional MSR approach, one employs a key approximation $\delta_n(s_h,Q^2_0) \simeq \delta_{n+1}(s_h,Q^2_0)$ for sufficiently large $n$, which serves as a sufficient condition. Under this assumption, the phenomenological representation retains only the ground-state contribution, whereas the OPE side continues to include contributions from higher excited states. Imposing the quark-hadron duality in Eq.~\eqref{eq:new duality}, which gives $\delta'_n(s_0,Q^2_0)=\delta_n(s_h,Q^2_0)$, it inevitably follows that $\delta'_n(s_0,Q^2_0) \simeq \delta'_{n+1}(s_0,Q^2_0)$. This condition is the foundation of the IMSR approach proposed in this work.

Incorporating the approximations yields the mass formula
\begin{equation}
	\label{eq:mass of improve moment}
	\begin{aligned}
		m_X^2(s_0,Q^2_0,n)=\frac{\tilde{M}_n^{\text{OPE}}(s_0,Q^2_0)}{\tilde{M}_{n+1}^{\text{OPE}}(s_0,Q^2_0)}-Q_0^2.
	\end{aligned}
\end{equation}
Compared with the conventional MSR in Eq.~\eqref{eq:mass of old moment}, the improved expression in Eq.~\eqref{eq:mass of improve moment} differs only in that the upper limit of integration in the moment function is now truncated at the scale $s_0$.

Unlike the unknown $\delta_n(s_h,Q^2_0)$, the quantity $\delta'_n(s_0,Q^2_0)$ can be determined by the OPE series.
Given that $\rho^{\text{OPE}}(s)$ is known, $\delta'_n(s_0,Q^2_0)$ depends on the free parameters $(s_0,Q^2_0,n)$. 
The approximation $\delta'_n(s_0,Q^2_0) \simeq \delta'_{n+1}(s_0,Q^2_0)$ thus provides an a posteriori constraint on $s_0$.

It should be further emphasized that $\delta_n(s_h,Q^2_0) \simeq \delta_{n+1}(s_h,Q^2_0)$ on the phenomenological side is not merely an approximation. In effect, the transition from Eq.~\eqref{eq:PH side of moment} to Eq.~\eqref{eq:easy PH side of moment} demands that $\delta_n(s_h,Q^2_0)$ and  $\delta_{n+1}(s_h,Q^2_0)$ be strictly equal. Consistency then requires $\delta'_n(s_0,Q^2_0) = \delta'_{n+1}(s_0,Q^2_0)$ on the OPE side as well.
In practice, however, the equality cannot be achieved exactly, as the parameters $(s_0,Q^2_0,n)$ are inherently finite in any numerical implementation. We will therefore enforce this condition in an approximate sense by means of a fitting procedure, which will be detailed in the following.


In the IMSR framework, we are able to calculate the coupling $f_X$ by using the quark-hadron duality 
\begin{equation}
	\label{eq:coupling in sum rules}
	\begin{aligned}
		\int_{s_N}^{s_0}\frac{\rho^{\text{OPE}}(s)}{s-q^2}\mathrm{d}s= \frac{f^2_X}{m^2_X-q^2}.
	\end{aligned}
\end{equation}
Taking the $n$-th moment of both sides 
\begin{equation}
	\begin{aligned}
		\tilde{M}^{\text{OPE}}_n(s_0,Q_0^2)=\frac{f^2_X}{(m^2_X+Q^2_0)^{n+1}},
	\end{aligned}
\end{equation}
so that the coupling $f_X$ can then be determined as
\begin{equation}
	\label{eq:coupling in improve moment}
	\begin{aligned}
		f^2_X=(m^2_X+Q^2_0)^{n+1}\tilde{M}^{\text{OPE}}_n(s_0,Q_0^2).
	\end{aligned} 
\end{equation}
It is clear that the coupling $f_X$ depends on the hadron mass $m_X$, and both of them must be evaluated with the same set of parameters $(s_0,Q^2_0,n)$.

\subsection{Dependence conditions}
The previous subsection introduced the approximation $\delta'_n(s_0,Q^2_0) = \delta'_{n+1}(s_0,Q^2_0)$, which is adopted to derive the mass formula in Eq.~\eqref{eq:mass of improve moment}. However, a single condition is clearly insufficient to determine the three parameters $(s_0,Q^2_0,n)$. Additional restrictions are therefore required.

In the LSR, the ``narrow resonance'' ansatz implies that the hadron mass should be independent of the Borel parameter. Motivated by the discussion in Ref.~\cite{Albuquerque:2026gll}, we examine the same ``narrow resonance'' ansatz applied in the IMSR. To this end, we define the normalized weighted spectral measure on $s \in [s_N,s_0]$
\begin{equation}
	\mathrm{d}\mu_n(s)=\frac{1}{\tilde{M}_n(s_0,Q^2_0)} \frac{\rho(s)}{(s+Q^2_0)^{n+1}}\mathrm{d}s,
	\label{eq:weighted function}
\end{equation}
and the expectation value for a function $f(s)$ 
\begin{equation}
	E_n[f]=\int f(s) \ \mathrm{d}\mu_n(s).
\end{equation}
Then the mass formula in Eq.~\eqref{eq:mass of improve moment} can be written as
\begin{equation}
	m^2_X+Q^2_0=\frac{\tilde{M}_n(s_0,Q^2_0)}{\tilde{M}_{n+1}(s_0,Q^2_0)}=E_n^{-1}[(s+Q^2_0)^{-1}], \label{eq:mass equation with tjx meaning}
\end{equation}
where $m^2_X+Q^2_0$ is actually the harmonic mean of $s+Q^2_0$ under the weighted spectral distribution. Differentiating with respect to $Q^2_0$ and $n$ one obtains
\begin{equation}
	\begin{aligned}
	\frac{\partial m_X^2}{\partial Q_0^2}
	&=
	(n+2)(m_X^2+Q_0^2)^2\,
	\mathrm{Var}_n\!\left[\frac{1}{s+Q_0^2}\right], \\
	\frac{\partial m_X^2}{\partial n}
	&=
	(m_X^2+Q_0^2)^2\,
	\mathrm{Cov}_n\!\left[\frac{1}{s+Q_0^2},\ln(s+Q_0^2)\right],\label{eq:differentiating with Q and n}
	\end{aligned}
\end{equation}
where the variance $\mathrm{Var}_n$ and covariance $\mathrm{Cov}_n$ are 
defined in Eq.~\eqref{variance} and Eq.~\eqref{covariance} respectively. 
The derivation of Eq.~\eqref{eq:differentiating with Q and n} is presented in Appendix~\ref{sec:dependence meaning}. These relations show that the dependence of $m^2_X$ on $Q^2_0$ and $n$ is governed by the variance and covariance of the weighted spectral distribution. If the spectral measure were an exact single $\delta$-function at $s=m^2_X$, both the variance and covariance would vanish identically, and $m_X$ would be independent of $Q^2_0$ and $n$. Conversely, any nonzero variance or covariance implies that the weighted spectral distribution has a finite width and therefore deviates from the single-$\delta$ ansatz adopted on the phenomenological side.

From the above discussions, the ``narrow resonance'' assumption on the phenomenological side requires that the extracted hadron mass should be stable against variations in both $Q^2_0$ and $n$. In practice, this requirement can be recast as conditions on the parametric dependence that the improved mass formula in Eq.~\eqref{eq:mass of improve moment} must satisfy:
\begin{widetext}
	\begin{equation}
		\label{eq:dependence condition}
		\begin{aligned}
			\left| \frac{\partial m^2_X(s_0,Q^2_0,n)}{\partial Q^2_0}  \right| & = \left| \frac{1}{\tilde{M}^2_{n+1}}\left( \tilde{M}_{n+1}\frac{\partial \tilde{M}_n}{\partial Q^2_0}-\tilde{M}_{n}\frac{\partial \tilde{M}_{n+1}}{\partial Q^2_0}-\tilde{M}^2_{n+1} \right) \right|   \leq \epsilon_Q ,\\
			\left| \frac{\partial m^2_X(s_0,Q^2_0,n)}{\partial n} \right| & = \left| \frac{1}{\tilde{M}^2_{n+1}}\left( \tilde{M}_{n+1}\frac{\partial \tilde{M}_n}{\partial n}-\tilde{M}_{n}\frac{\partial \tilde{M}_{n+1}}{\partial n} \right) \right| \leq \epsilon_n \ \mathrm{GeV}^2,
		\end{aligned}
	\end{equation}
\end{widetext}
in which $\epsilon_Q$ and $\epsilon_n$ are two small tolerance parameters. The practical importance of Eq.~\eqref{eq:dependence condition} lies in its ability to provide a quantitative measure of parametric dependences, thereby offering a well-defined criterion for assessing the reliability of the extracted mass against variations in the auxiliary parameters. Assuming that $Q^2_0$ and $n$ similarly affect the mass $m_X$, we shall adopt $\epsilon_Q=\epsilon_n=\epsilon$ in the following section for simplicity and without loss of generality.
%
The dimensionless $\epsilon$ quantifies the allowed deviation from the ideal single-pole limit, i.e. the degree of concentration of the weighted spectral distribution. 

It should be emphasized that these conditions on the parametric dependence are necessary but not sufficient. A vanishing derivative does not by itself prove that the spectral measure is a single $\delta$-function; a large derivative, however, unambiguously signals that the single-$\delta$ ansatz is not self-consistent. In this sense, the conditions provide a well-defined necessary criterion for identifying the parameter region in which the ``narrow resonance'' approximation is reliable.

These conditions are therefore essentially self-consistency conditions of the model. They require that the ``narrow resonance'' assumption be satisfied as closely as possible on the OPE side, since this is precisely the model employed on the phenomenological side.

For further elaboration, we note that the approximation condition $\delta'_n(s_0,Q^2_0) \simeq \delta'_{n+1}(s_0,Q^2_0)$ is intrinsically guaranteed by the quark-hadron duality in the derivation of Eq.~\eqref{eq:mass of improve moment}. Meanwhile, the conditions in Eq.~\eqref{eq:dependence condition} are also essentially self-consistency conditions of the model: they require that the OPE side agree with the phenomenological side as closely as possible. Consequently, the procedure of extracting physical results by imposing these constraints closely mirrors the Bootstrap philosophy, in which key physical quantities are determined through self-consistency conditions and physical constraints. We emphasize that the quark-hadron duality is fundamentally a phenomenological assumption whose validity must ultimately be tested and verified against experimental data~\cite{Wunderlich:2026krs,Bloom:1970xb,Bloom:1971ye,Melnitchouk:2005zr,JeffersonLabE00-115:2009jll}, so that the methodology introduced here represents only a generalized Bootstrap-like approach.
	

\subsection{General procedure}
In this subsection, we demonstrate how the approximation and dependence conditions enable a unique determination of both $s_0$ and $m_X$, without invoking any subjective criteria.

We begin by choosing a trial interval $[n_0,n_1]$ for the integer parameter $n$. For fixed $s_0$ and $n$, the conditions in Eq.~\eqref{eq:dependence condition} constrain $Q_0^2$ to a finite range, so that no a priori interval is needed for this variable. In the following, we discretize $Q^2_0$ with a grid spacing of $0.5 \ \mathrm{GeV}^2$ for the numerical analysis.

We next select a value of $\epsilon$ and investigate, for a range of $s_0$ values, whether Eq.~\eqref{eq:dependence condition} admits solutions for $Q^2_0$ within the chosen interval $ n \in [n_0,n_1]$. When $s_0$ is significantly removed from its physical value --- for instance, $s_0=2$ or $8\ \mathrm{GeV}^2$, as opposed to the expected vicinity of $5\ \mathrm{GeV}^2$ --- the conditions in Eq.~\eqref{eq:dependence condition} are generally not satisfied. This behavior provides a practical means of identifying the approximate range in which $s_0$ must reside.

Decreasing $\epsilon$ imposes progressively stricter parametric dependence, which in turn narrows the viable interval for $s_0$. This is advantageous: it reduces the computational cost of subsequent analyses and simultaneously removes unphysical regions from the parameter space. To determine a suitable $\epsilon$, we adopt an iterative strategy, beginning with a relatively loose threshold (e.g., $\epsilon=10^{-3}$) and gradually tightening it (e.g., to $\epsilon=10^{-4}$), while observing the corresponding contraction of the acceptable $s_0$ region.

To quantify the approximation, we define the function $\eta(s_0,Q^2_0,n)$ as
\begin{equation}
	\label{eq:approximate function}
	\begin{aligned}
		\eta(s_0,Q^2_0,n)=\frac{\delta'_n(s_0,Q_0^2)}{\delta'_{n+1}(s_0,Q_0^2)}.
	\end{aligned}
\end{equation}
As discussed above, deriving Eq.~\eqref{eq:mass of improve moment} from Eq.~\eqref{eq:easy equ in new moment} requires that
\begin{equation}
	\begin{aligned}
		\eta(s_0,Q^2_0,n)=1,
	\end{aligned}
\end{equation}
which can be realized only in the limit $n\rightarrow \infty$, i.e.
\begin{equation}
	\begin{aligned}
		\lim_{n \to \infty} \frac{\delta'_n(s_0,Q_0^2)}{\delta'_{n+1}(s_0,Q_0^2)} =1.
	\end{aligned}
\end{equation}
Note that $\eta(s_0, Q^2_0,n)$ is strictly monotonically decreasing with respect to $n$, which implies $\eta>1$ for all finite values of $n>0$. The proof of this key property is deferred to Appendix~\ref{sec:eta}. This monotonic behavior is precisely the reason why we choose to scan over a trial range of $n$ rather than over $Q^2_0$ in our numerical procedure.

After determining the parameter space $(s_0, Q^2_0,n)$ that satisfies the dependence conditions in Eq.~\eqref{eq:dependence condition} within a given   interval $ [n_0, n_1]$, we locate the minimum of $\eta(s_0, Q^2_0,n)$ and  denote it by $\eta_{\text{min}}$, together with the value of $s_0$. Scanning over successive intervals $ [n_i,n_{i+1}]$ for $i=1,2,3,\ldots$, we build up the sequence $(\eta^i_{\text{min}}, s_0^i)$. Because $n$ is restricted to a finite set of integers in each interval, $\eta_{\text{min}}$ remains strictly greater than unity. We then perform a fit of $s_0$ against $\eta_{\text{min}}$ and extrapolate the resulting curve to the limiting point $\eta_{\text{min}}=1$, thereby determining the limiting value of $s_0$ that corresponds to the ideal limit of the approximation.

Further details of this procedure will be elaborated in the applications presented in Sec.~\ref{sec:numerical analysis}.

%
%
%
%

\subsection{The $n \rightarrow \infty$ limit}\label{sec:two logical routes}

One may wonder why the \(n\to\infty\) limit is necessary. From a purely mathematical standpoint, substituting quark-hadron duality directly into Eq.~\eqref{eq:easy equ in new moment} yields Eq.~\eqref{eq:mass of improve moment} without requiring \(n\to\infty\). Here we explain the rationale for this choice.

Starting from Eq.~\eqref{eq:easy equ in new moment}, there are two logically consistent ways to arrive at the improved mass formula in Eq.~\eqref{eq:mass of improve moment}.

\emph{Route 1:} Using the quark-hadron duality \(\delta'_n(s_0,Q_0^2)=\delta_n(s_h,Q_0^2)\), the two ratios in Eq.~\eqref{eq:easy equ in new moment} are equal and cancel each other for any finite \(n\), and then one can obtain Eq.~\eqref{eq:mass of improve moment}.

\emph{Route 2:} Taking the \(n\to\infty\) limit ensures that \(\delta_n=\delta_{n+1}\) and \(\delta'_n=\delta'_{n+1}\). Both ratios approach unity, and Eq.~\eqref{eq:mass of improve moment} follows.

\begin{figure*}[htbp]
	\centering
	\subfigure[]{
		\includegraphics[width=0.3\textwidth]{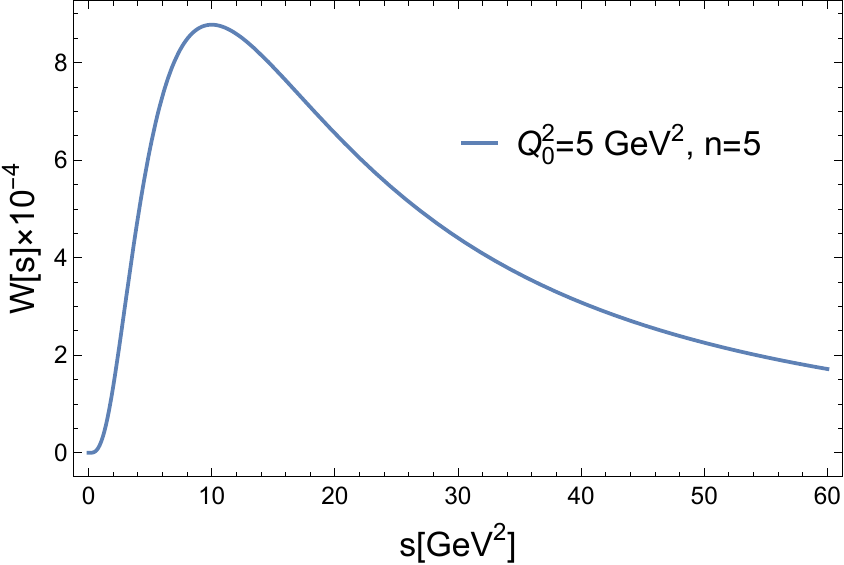}
	}
	\subfigure[]{
		\includegraphics[width=0.3\textwidth]{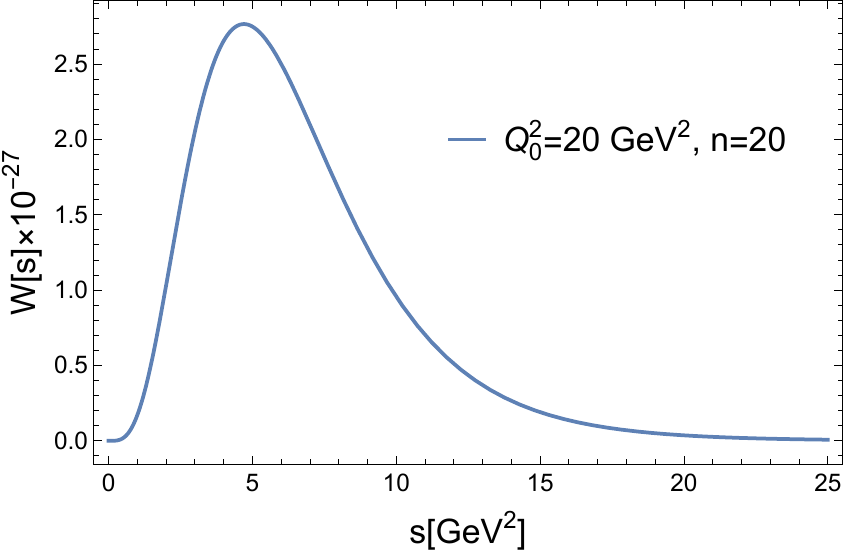}
	}
	\subfigure[]{
		\includegraphics[width=0.3\textwidth]{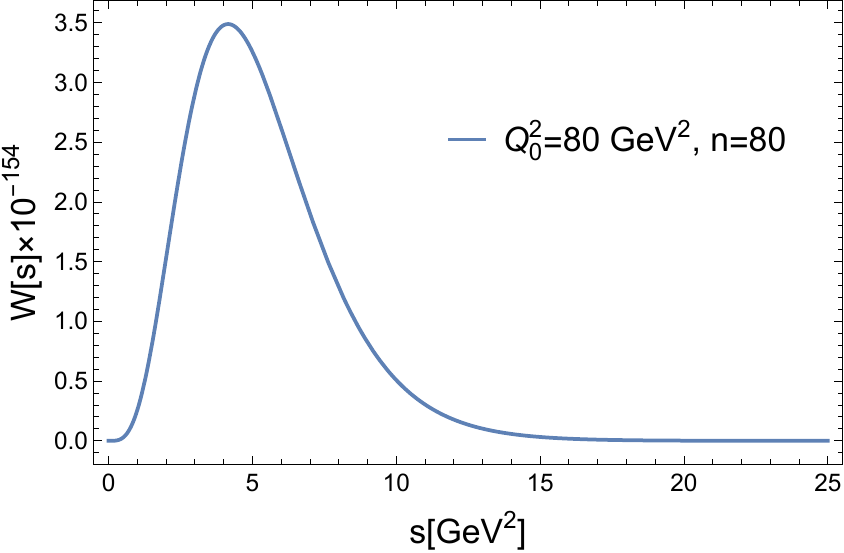}
	}
	\caption{Illustration of the narrowing effect of the weighted spectral function $W(s)=\frac{s^4}{(s+Q^2_0)^{n+1}}$ as $n$ increases, with the ratio $Q^2_0/n$ fixed at $1\ \mathrm{GeV}^2$. Panels (a), (b), and (c) correspond to $n=5$, $20$, and $80$, respectively.}\label{fig:wighted function with different n}
\end{figure*}

We adopt Route 2 to achieve the mass formula. The motivation, however, is not merely to obtain an internal constraint on the parameters. More fundamentally, the phenomenological side uses the single-$\delta$ ``narrow resonance'' assumption, which corresponds to a zero-width spectral distribution. As shown in Fig.~\ref{fig:wighted function with different n}, the weighted spectral distribution will be suppressed to the low-energy region by the weight factor \((s+Q_0^2)^{-n-1}\) for the increasing $n$ and a fixed ratio $Q^2_0/n$. The larger \(n\) is, the smaller the width of the distribution becomes, and the better it matches the zero-width ``narrow resonance'' ansatz on the phenomenological side. In other words, the \(n\to\infty\) limit is adopted to minimize the systematic error of the calculation, not merely to introduce an internal constraint.

It should also be noted that even in the limit \(n\to\infty\), the width of the spectral distribution on the OPE side does not tend to zero; it is proportional to \(Q_0^2/n\) --- a ratio that, as we will see later, is constrained by the dependence conditions to be approximately constant, mirroring \(M_B^2\) in the LSR. Therefore, this systematic error is unavoidable and can only be reduced, not eliminated. This should be understood as a limitation of the ``narrow resonance'' approximation as a phenomenological model.

\section{Numerical analysis and applications}\label{sec:numerical analysis}
\subsection{A light tetraquark state}
In this subsection, we illustrate the IMSR by applying it to a pseudoscalar light tetraquark $ud\bar{d}\bar{s}$ system. Specifically, we employ the following interpolating current with $J^P=0^-$ introduced in Ref.~\cite{Zhang:2025fuz} 
\begin{equation}
	P_6=u_a^T \mathcal{C}  \gamma_\mu \gamma_5 d_b [\bar{d}_a \gamma_\mu \mathcal{C} \bar{s}^T_b + (a \leftrightarrow b)],
\end{equation}
where $u, d$, and $s$ are up, down and strange quark fields, respectively, $a$ and $b$ are color indices, and $\mathcal{C}= \mathrm{i} \gamma_2 \gamma_0$ is the charge conjugation operator. The details of the OPE calculations were given in Ref.~\cite{Zhang:2025fuz}, where the spectral density up to dimension-8 reads
\begin{widetext}
	\begin{equation}
		\label{eq:OPE of P6}
		\begin{aligned}
			& \rho_{P_6}^{\text{OPE}}(s)  =   \Bigg[  \frac{s^4}{2^9 3^1 5\pi^6} -  \left( \frac{m_s \langle \bar{s}s \rangle}{2^4 3\pi^4}-\frac{m_s \langle \bar{q}q \rangle}{2^4 3\pi^4} +\frac{5 \langle g^2 G^2 \rangle}{2^{10} 3\pi^6} \right)  s^2 - \Bigg(  \frac{7 g^2 \langle \overline{s} s \rangle^2}{2^4 3^4
				\pi^4} + \frac{7 g^2 \langle \overline{q} q
				\rangle^2}{2^4 3^3 \pi^4}\\ 
			&   - \frac{113 \langle g^3 f G^3 \rangle}{2^9 3^3 \pi^6} - \frac{m_s
				\langle g \overline{q}  \sigma G q \rangle}{2^7 \pi^4} + \frac{7 m_s \langle g
				\overline{s} \sigma G s \rangle}{2^7 3 \pi^4} - \frac{\langle \overline{q} q
				\rangle^2}{6 \pi^2} + \frac{\langle \overline{q} q \rangle \langle
				\overline{s} s \rangle}{6 \pi^2}   \Bigg) s   \\
			& -\left(  \frac{5 m_s \langle g^2 G^2 \rangle \langle \overline{s} s \rangle}{2^8 3
				\pi^4}  - \frac{7 m_s \langle g^2 G^2 \rangle \langle \overline{q} q
				\rangle}{2^7 3^2 \pi^4} - \frac{\langle \overline{q} q \rangle \langle g
				\overline{q} \sigma G q \rangle}{2^3 3 \pi^2} + \frac{\langle g \overline{s}
				\sigma G s \rangle \langle \overline{q} q \rangle}{2^4 3 \pi^2} + \frac{\langle
				g \overline{q} \sigma G q \rangle \langle \overline{s} s \rangle}{2^4 3 \pi^2}
			\right) \Bigg] ,
		\end{aligned}
	\end{equation}
\end{widetext}
where $m_s$ denotes the strange quark mass.

For the numerical analysis, we employ the following set of input parameters, including the quark masses, the strong coupling constant, and the QCD condensates~\cite{ParticleDataGroup:2024cfk,Narison:2018dcr,Narison:2025cys,Jin:2002rw,Huang:2014hya,DiGiacomo:1982gn,Shifman:1978by}:
\begin{equation}
	\label{eq:value of condensate}
	\begin{gathered}
		m_u=m_d=m_q =  0, \  m_s(2\ \mathrm{GeV})=93.5\pm 0.8\ \mathrm{MeV}, \\
		\ \langle \bar{q}q \rangle =-(0.24\pm0.01)^3 \ \mathrm{GeV}^3 ,    \langle \bar{s}s \rangle  =(0.8\pm0.1)\langle \bar{q}q \rangle , \\
		\langle  g \bar{q} \sigma G q  \rangle =(0.8\pm0.2) \langle \bar{q}q \rangle \ \mathrm{GeV}^2, \\
		\langle g \bar{s} \sigma G s \rangle =(0.8\pm0.2) \langle g \bar{q} \sigma G q \rangle,
		\\
		\langle \alpha_s  G^2 \rangle  = (6.35\pm0.35) \times 10^{-2} \ \mathrm{GeV}^4, \\ \langle g^3 f G^3 \rangle =1.2 \mathrm{GeV}^2 \langle \alpha_s G^2 \rangle , \\
		m_s(\mu)=m_s(2\ \mathrm{GeV}) \Bigg[ \frac{\alpha_s(\mu)}{\alpha_s(2\ \mathrm{GeV})}\Bigg]^{\frac{12}{33-2n_f}}, \\  \alpha_s=\frac{1}{b_0 t}\Bigg[  1-\frac{b_1}{b_0^2} \frac{\mathrm{ln}t}{t} +\frac{b_1^2(\mathrm{ln}^2t-\mathrm{ln}t-1)+b_0 b_2}{b_0^4t^2} \Bigg],
	\end{gathered}
\end{equation}
where $t = \log \frac{\mu^2}{\Lambda^2}$, $b_0 = \frac{33 - 2 n_f}{12 \pi}$,
$b_1 = \frac{153 - 19 n_f}{24 \pi^2}$, $b_2 = \frac{2857 - \frac{5033}{9} n_f
	+ \frac{325}{27} n_f^2}{128 \pi^3}$, $\Lambda = 332 \,\text{MeV}$ for the number of quark 
flavors $n_f = 3$. The strange quark mass is taken in the
$\overline{\text{MS}}$ scheme at the scale $\mu = 2 \,\text{GeV}$. Given that
condensate values are evaluated at the renormalization scale $\mu = 1\ \mathrm{GeV}$, we
accordingly adopt the quark masses at the same scale.

\begin{figure*}[htbp]
	\centering
	\subfigure[]{
		\includegraphics[width=0.3\textwidth]{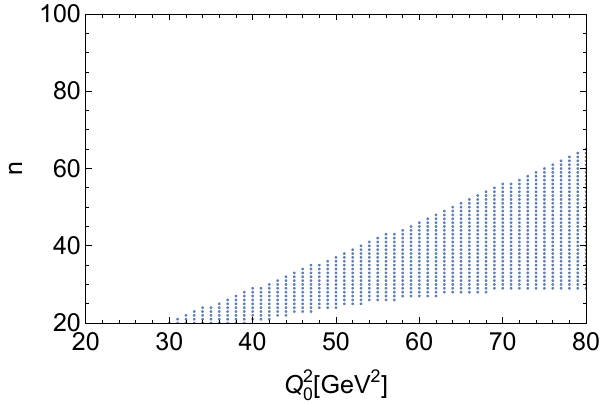}
	}
	\subfigure[]{
		\includegraphics[width=0.3\textwidth]{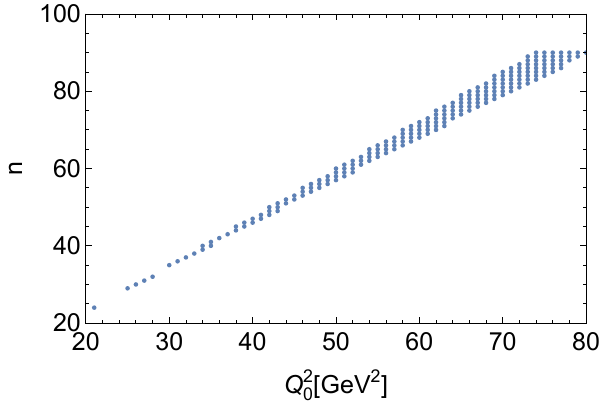}
	}
	\\
	\subfigure[]{
		\includegraphics[width=0.3\textwidth]{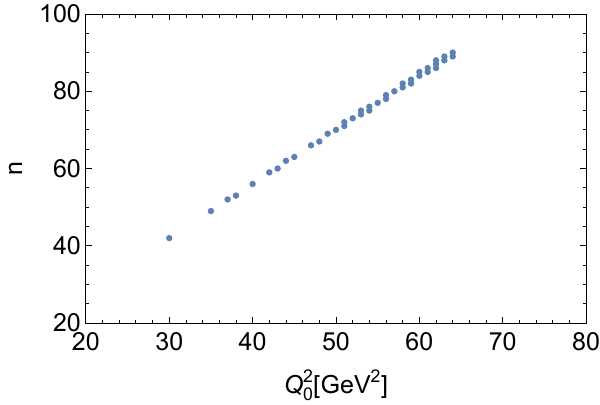}
	}
	\subfigure[]{
		\includegraphics[width=0.3\textwidth]{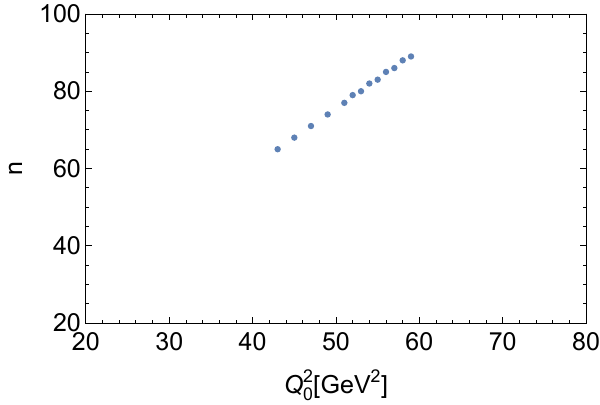}
	}
	\subfigure[]{
		\includegraphics[width=0.3\textwidth]{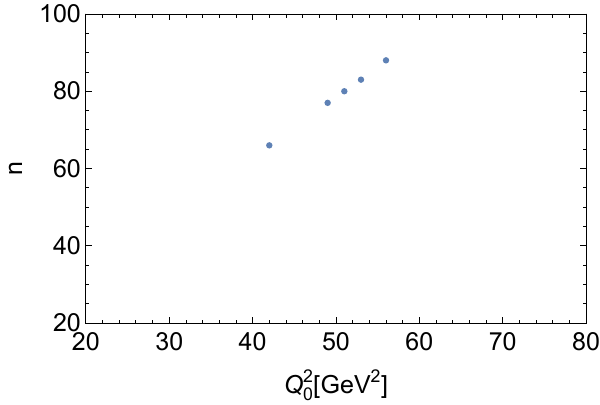}
		\label{fig:test for dependence with s=3}
	} 
	\caption{Numerical test of the dependence condition in Eq.~\eqref{eq:dependence condition} for the OPE expression in Eq.~\eqref{eq:OPE of P6}, with $\epsilon=10^{-3}$. Panels (a)–(e) correspond to $s_0=3, 4, 5, 6, 7\ \mathrm{GeV}^2$ , respectively. The marked positions indicate parameter sets that satisfy the dependence condition in Eq.~\eqref{eq:dependence condition}.}
	\label{fig:test for dependence}
\end{figure*}

As an initial step, we select a trial range for $s_0$ and test whether the dependence conditions in Eq.~\eqref{eq:dependence condition} can be satisfied. Specifically, we consider $s_0=3,4,5,6,7\ \mathrm{GeV}^2$ with $\epsilon=10^{-3}$ , and the corresponding results are shown in Fig.~\ref{fig:test for dependence}. A more stringent value of $\epsilon$ can be adopted to further constrain $s_0$. With $\epsilon=0.25\times 10^{-3}$, we find that the conditions cannot be fulfilled for any $(Q^2_0,n)$ when $s_0>5 \ \mathrm{GeV}^2$.

\begin{figure*}[htbp]
	\centering
	\includegraphics[width=.5\textwidth]{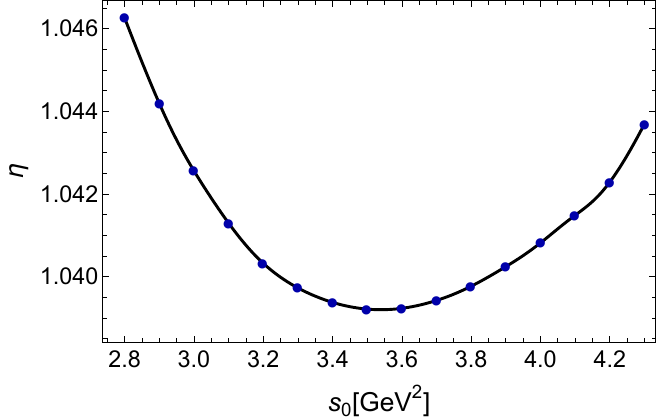}
	\caption{Variation of $\eta(s_0,Q^2_0,n)$ with $s_0$ for current $P_6$, with $\epsilon=0.25\times 10^{-3}$ and $n \in [60,80]$.}
	\label{fig:eta of s0}
\end{figure*}

\begin{figure*}[htbp]
	\centering
	\subfigure[]{
		\includegraphics[width=0.3\textwidth]{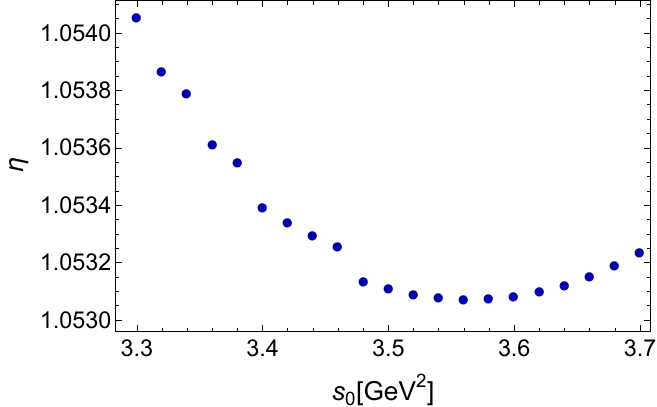}
	}
	\subfigure[]{
		\includegraphics[width=0.3\textwidth]{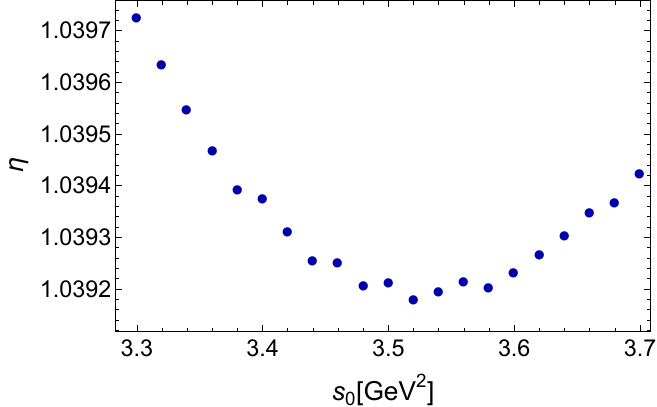}
		\label{fig:etamin versus s0 with 6080}
	}
	\subfigure[]{
		\includegraphics[width=0.3\textwidth]{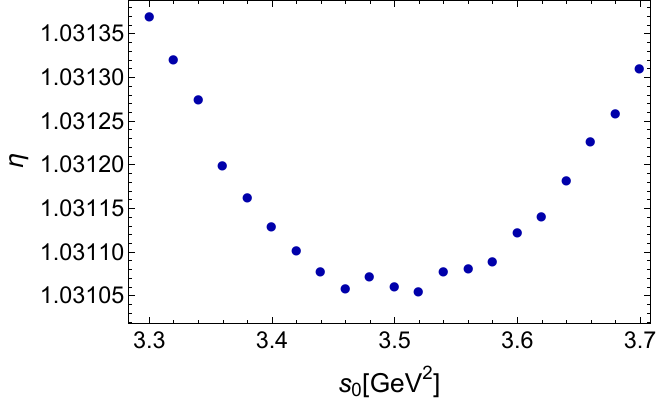}
	}
	\\
	\subfigure[]{
		\includegraphics[width=0.3\textwidth]{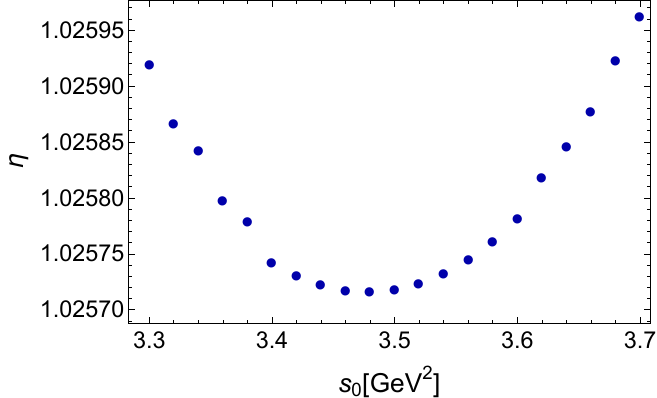}
	}
	\subfigure[]{
		\includegraphics[width=0.3\textwidth]{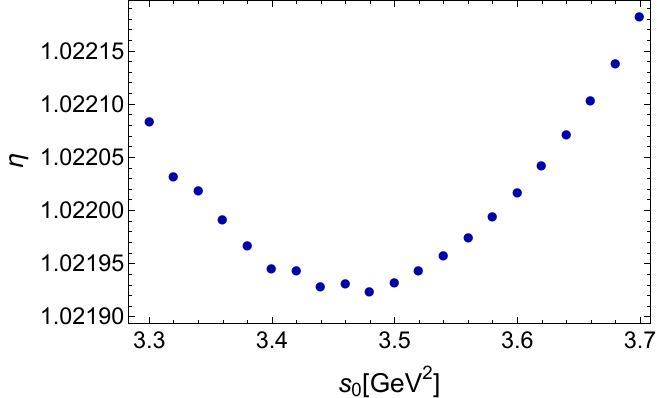}
	}
	\caption{Variations of $\eta(s_0,Q^2_0,n)$ with $s_0$ for current $P_6$ with $\epsilon=0.25\times 10^{-3}$ under different $n$-intervals: (a) $n \in [40,60]$, (b) $[60,80]$, (c) $[80,100]$, (d) $[100,120]$, (e) $[120,140]$.}
	\label{fig:etamin versus s0 with all easy}
\end{figure*}

For $\epsilon=0.25\times 10^{-3}$ and $n \in [60,80]$, we perform a scan over $s_0$ with a step size of $0.1\ \mathrm{GeV}^2$, and examine the optimal approximation value $\eta_{\text{min}}$ achievable within the parameter space $(s_0,Q^2_0,n)$. The results are presented in Fig.~\ref{fig:eta of s0}, where a clear minimum is observed at $(s_0^\star,\eta_{\text{min}})$. In other words, the procedure amounts to finding the global minimum of the function $\eta(s_0,Q^2_0,n)$ over the three-dimensional parameter space $(s_0,Q^2_0,n)$, while simultaneously requiring that the dependence conditions in Eq.~\eqref{eq:dependence condition} are satisfied.

The initial scan yields a minimum $\eta_{\text{min}}=1.03921$  at $s_0^\star=3.5\ \mathrm{GeV}^2$. Since the scan step size is $0.1\ \mathrm{GeV}^2$, this naturally translates to an uncertainty $s_0^\star=3.5\pm0.1\ \mathrm{GeV}^2$, avoiding any subjective error estimate. To improve the precision, we then perform a refined scan over the interval $s_0 \in [3.3,3.7]\ \mathrm{GeV}^2$ with a reduced step size of $0.02\ \mathrm{GeV}^2$. The refined result, shown in Fig.~\ref{fig:etamin versus s0 with 6080}, gives $s_0^\star=3.52\pm0.02\ \mathrm{GeV}^2$ and $\eta_{\text{min}}=1.03918$.

To assess the sensitivity of $(s_0^\star,\eta_{\text{min}})$ to the choice of the $n$-intervals, we repeat the same analysis for several alternative ranges: $n \in [40,60], [80,100], [100,120]$, and $[120,140]$. The resulting curves are displayed in Fig.~\ref{fig:etamin versus s0 with all easy}. The slight irregularity observed in the data points is attributable to the discrete nature of the $Q^2_0$ grid.

\begin{figure*}[htbp]
	\centering
	\includegraphics[width=0.5\textwidth]{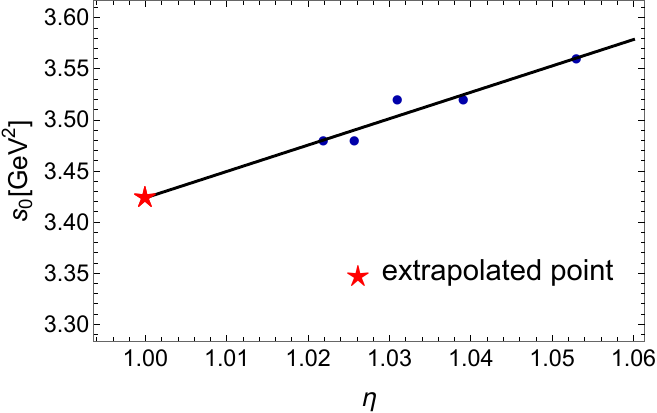}
	\caption{Extracted minima $(s_0^\star,\eta_{\text{min}})$ from Fig.~\ref{fig:etamin versus s0 with all easy}. The plot illustrates the variation of the minimum position $s^\star_0$ and its corresponding $\eta_{\text{min}}$ as the $n$-interval changes.}
	\label{fig:eta_min_s0_with_different_n1}
\end{figure*}

The results in Fig.~\ref{fig:etamin versus s0 with all easy} demonstrate that increasing the $n$-interval systematically reduces $\eta_{\text{min}}$ while simultaneously shifting the location of  $s_0^\star$ toward smaller values. By extracting the minimum coordinates $(s_0^\star,\eta_{\text{min}})$ from each curve, we obtain the data points plotted in Fig.~\ref{fig:eta_min_s0_with_different_n1}. This figure clearly shows that a larger $n$-interval corresponds to smaller values of both $\eta_{\text{min}}$ and $s_0^\star$. To determine the value of $s_0$ at the ideal limit $\eta=1$, we perform a linear fit to the data points and extrapolate, yielding
\begin{equation}
	\begin{aligned}
		s_0(\eta=1) = 3.42 \pm 0.03  \ \mathrm{GeV}^2, \label{Ks0}
	\end{aligned}
\end{equation}
where the uncertainty includes the $s_0$ step size and the linear-fit extrapolation error. Substituting this value into Eqs.~\eqref{eq:mass of improve moment} and \eqref{eq:coupling in improve moment} yields the hadron mass and the squared coupling
\begin{equation}
	\begin{aligned} 
		m_X =1.63 \pm 0.10 \ \mathrm{GeV}, \quad  f^2_X=(2.31 \pm 0.34)\times 10^{-5} \ \mathrm{GeV}^{10},
		\label{Kmass}
	\end{aligned}
\end{equation}
where the error includes both $s_0$ and condensate uncertainties. The extrapolation is performed over the full $n$-interval $[40,140]$, with a step size of $\Delta n=20$, $\Delta s_0=0.02 \ \mathrm{GeV}^2$, and $\epsilon=0.25\times 10^{-3}$.

\begin{figure*}[htbp]
	\centering
	\subfigure[]{
		\includegraphics[width=0.45\textwidth]{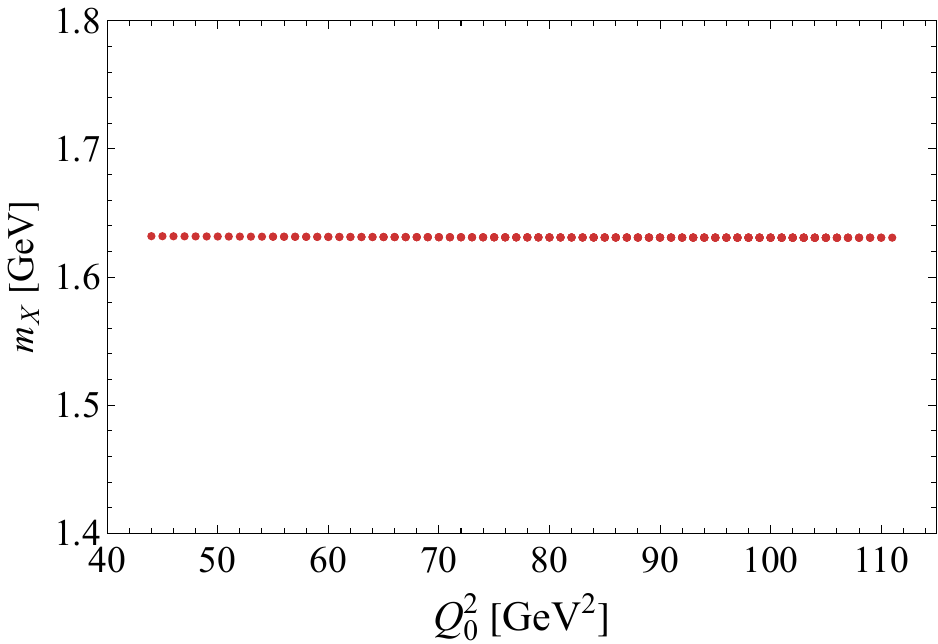}
	}
	\subfigure[]{
		\includegraphics[width=0.45\textwidth]{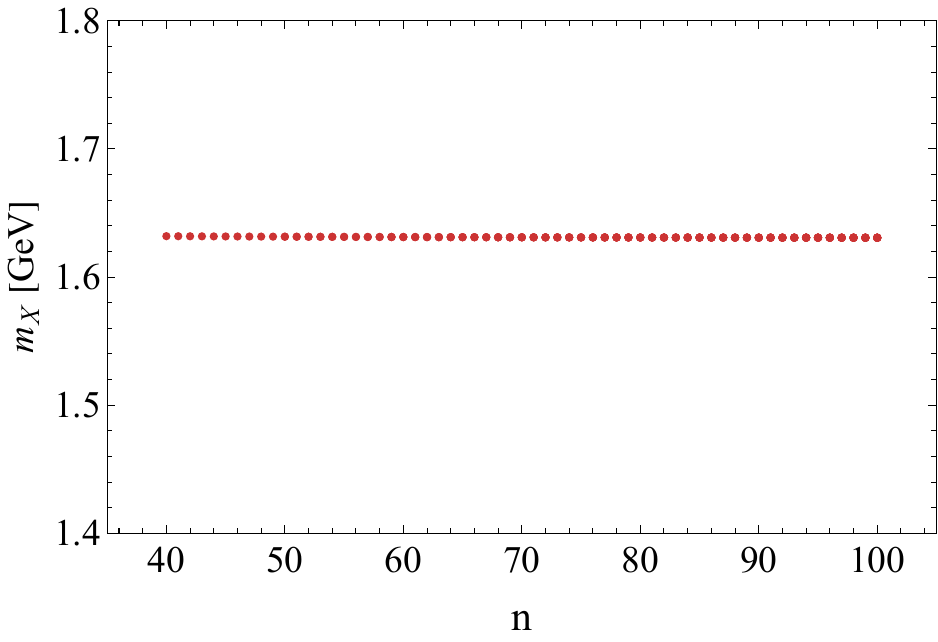}
	}
	\caption{Stability of the extracted mass $m_X$ against the parameters $Q^2_0$ (a) and $n$ (b). The parameters are restricted by the dependence conditions Eq.~\eqref{eq:dependence condition} at $s_0=3.42\ \mathrm{GeV}^2$.}\label{fig:1690 mass dependence}
\end{figure*}

\begin{figure*}[htbp]
	\centering
	\subfigure[]{
		\includegraphics[width=0.45\textwidth]{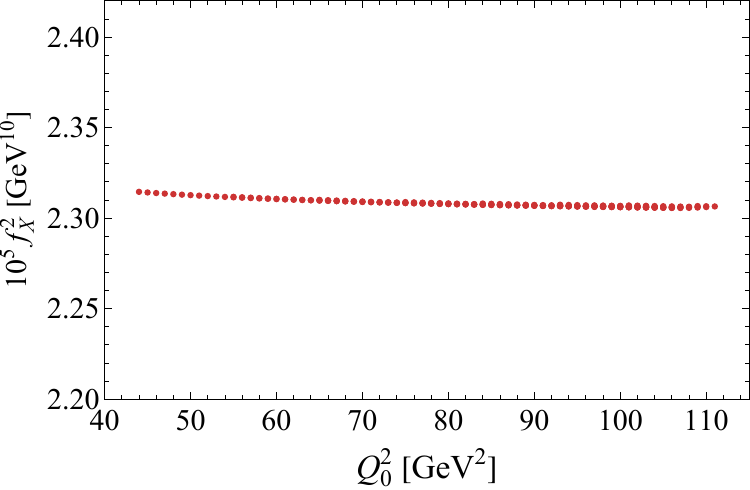}
	}
	\subfigure[]{
		\includegraphics[width=0.45\textwidth]{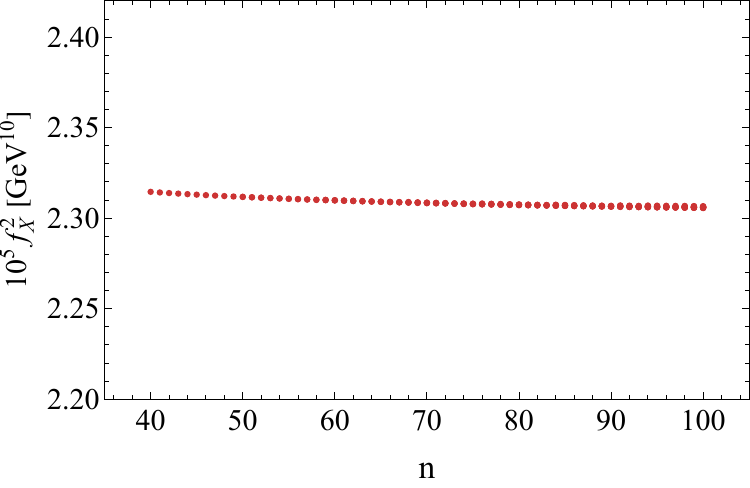}
	}
	\caption{Stability of the extracted coupling $f^2_X$ against the parameters $Q^2_0$ (a) and $n$ (b). The parameters are restricted by the dependence conditions Eq.~\eqref{eq:dependence condition} at $s_0=3.42\ \mathrm{GeV}^2$.}\label{fig:coupling dependence}
\end{figure*}
The extracted mass and coupling are presented in Figs.~\ref{fig:1690 mass dependence} and \ref{fig:coupling dependence}, respectively. The mass shows excellent stability against the parameters $Q^2_0$ and $n$. 
Notably, although the parameter space $(Q^2_0,n)$ is determined solely by imposing the mass-stability conditions in Eq.~\eqref{eq:dependence condition}, the extracted coupling $f^2_X$ is found to be similarly stable within this same parameter region, as illustrated in Fig.~\ref{fig:coupling dependence}. This simultaneous stability strongly demonstrates the robustness of the IMSR framework.

\begin{figure*}[htbp]
	\centering
	\subfigure[]{
		\includegraphics[width=0.45\textwidth]{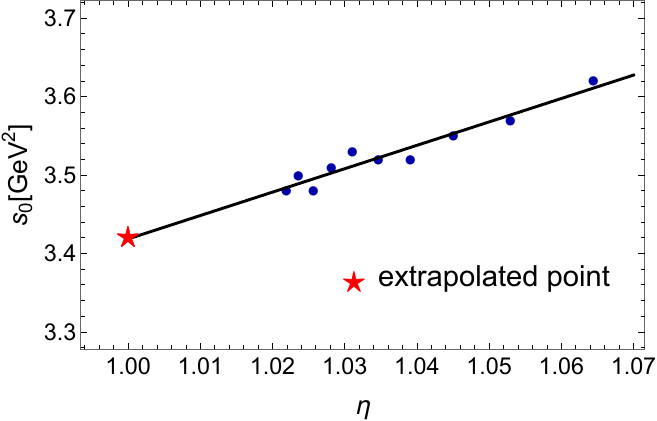}
	}
	\subfigure[]{
		\includegraphics[width=0.45\textwidth]{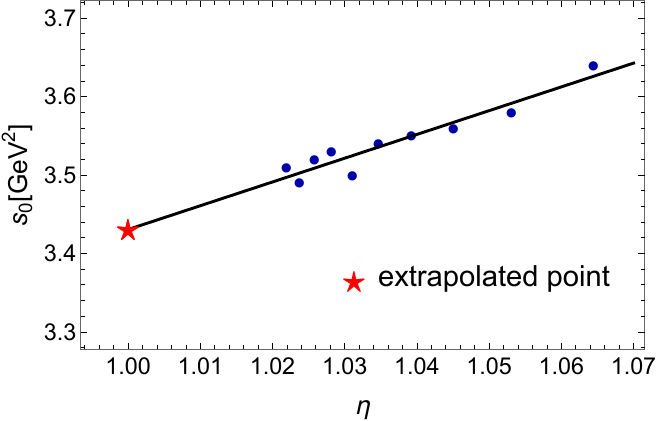}
	}
	\caption{Extrapolation results for current $P_6$ using $\Delta n=10$, $\Delta s_0=0.01 \ \mathrm{GeV}^2$ for $\epsilon=0.25\times 10^{-3}$ (a) and $\epsilon=10^{-4}$ (b). The asterisk indicates the extrapolated values at $\eta=1$.}\label{fig:eta_min_s0_with_different_n23}
\end{figure*}

To verify the robustness of the linear extrapolation, we repeat the entire procedure under two alternative configurations with finer step sizes: (i) $\Delta n=10$, $\Delta s_0=0.01 \ \mathrm{GeV}^2$, $\epsilon=0.25\times 10^{-3}$; and (ii) $\Delta n=10$, $\Delta s_0=0.01 \ \mathrm{GeV}^2$, $\epsilon=10^{-4}$. The results of these refined analyses are shown in Fig.~\ref{fig:eta_min_s0_with_different_n23}, yielding the extrapolated values $s_0(\eta=1)=3.42 \pm 0.02 \ \mathrm{GeV}^2$ and $s_0(\eta=1)=3.43 \pm 0.02 \ \mathrm{GeV}^2$, respectively.  These values agree well with the previous result, providing strong evidence for the stability and robustness of the extrapolation method. Importantly, the entire determination is substantially free of any ad hoc or subjective assumptions, in contrast to conventional sum rule analyses.

\subsection{The $\Sigma\bar{\Sigma}$ molecular state}

In this subsection, we apply the IMSR to the $\Sigma\bar{\Sigma}$ molecular state with $I(J^{PC})=0(0^{-+})$, which has been proposed as a possible interpretation of the $X(2370)$~\cite{Ben:2026afy}.

\begin{figure*}[htbp]
	\centering
	\subfigure[]{
		\includegraphics[width=0.2\textwidth]{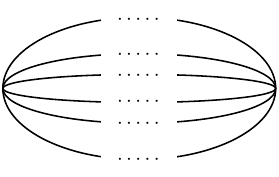}
	}
	\subfigure[]{
		\includegraphics[width=0.2\textwidth]{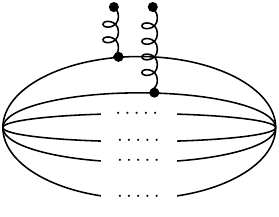}
	}
	\subfigure[]{
		\includegraphics[width=0.2\textwidth]{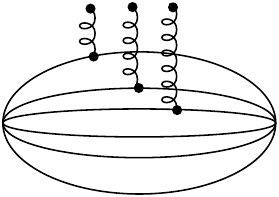}
	}
	\\
	\subfigure[]{
		\includegraphics[width=0.2\textwidth]{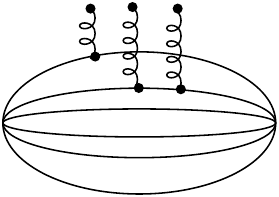}
	}
	\subfigure[]{
		\includegraphics[width=0.2\textwidth]{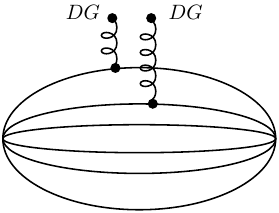}
	}
	\subfigure[]{
		\includegraphics[width=0.2\textwidth]{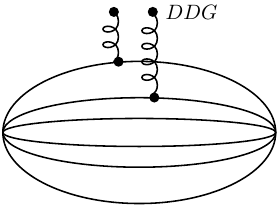}
	}
	\\
	\subfigure[]{
		\includegraphics[width=0.2\textwidth]{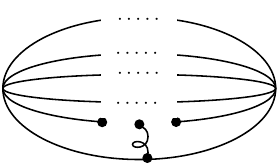}
	}
	\subfigure[]{
		\includegraphics[width=0.2\textwidth]{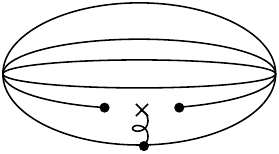}
	}
	\subfigure[]{
		\includegraphics[width=0.2\textwidth]{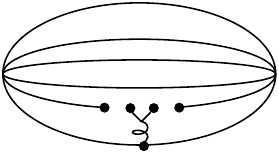}
	}
	\caption{The Feynman diagrams involved in our calculation for the $\Sigma\bar{\Sigma}$ molecular state. A quark line with dots contains terms with the color factor $\delta^{ij}$ in the quark propagator.}\label{fig:feynman}
\end{figure*}
We adopt $J_{\Sigma}=\epsilon^{abc}(q_a^T\mathcal{C} \gamma_\mu q_b)\gamma_5 \gamma_\mu q_c$ as the interpolating current for $\Sigma$~\cite{Ioffe:2010zz}, where $q$ denotes the quark field of flavor $u$, $d$, or $s$. 
Taking into account the Clebsch-Gordan coefficients, we can construct the interpolating current for the $\Sigma \bar{\Sigma}$ molecular state with $I=0$
\begin{align}
		J_{\bar{\Sigma}\Sigma}=J_{\bar{\Sigma}^+}\gamma_5 J_{\Sigma^+}+& J_{\bar{\Sigma}^0} \gamma_5  J_{\Sigma^0}+J_{\bar{\Sigma}^-}\gamma_5 J_{\Sigma^-},
\end{align}
where 
\begin{equation}
	\begin{aligned}
		J_{\Sigma^+}=\epsilon_{abc}(u_a^T\mathcal{C} \gamma_\mu u_b)\gamma_5 \gamma_\mu s_c &, \quad J_{\Sigma^-}=\epsilon_{abc}(d_a^T\mathcal{C} \gamma_\mu d_b)\gamma_5 \gamma_\mu s_c , 
		\\
		J_{\Sigma^0}=\frac{1}{\sqrt{2}}\epsilon_{abc}[(u_a^T\mathcal{C} \gamma_\mu d_b) & \gamma_5  \gamma_\mu  s_c+(d_a^T\mathcal{C} \gamma_\mu u_b)\gamma_5 \gamma_\mu s_c].
	\end{aligned}
\end{equation}

To calculate the correlation function, we consider the Feynman diagrams depicted in Fig.~\ref{fig:feynman} and adopt the full propagator as in Ref.~\cite{Zhang:2025fuz}. The OPE-side spectral density up to dimension-8 is then obtained as
\begin{widetext}
	\begin{equation}
		\begin{aligned}
			\rho_1^{\Sigma \bar{\Sigma}}(s) = & \frac{s^7}{2^{17}3^1 5^2 7^2 \pi^{10}} + \frac{m_s^2 s^6}{2^{16}3^1 5^2 7 \pi^{10}} + \left( \frac{\langle g^2 G^2 \rangle}{2^{16} 3^1 5^2 \pi^{10}} + \frac{m_s \langle \bar{s} s \rangle}{2^{12}3^1 5^2 \pi^8} \right) s^5 + \left( \frac{g^2 \langle \bar{s}s \rangle^2}{2^{9}3^4 5\pi^8} - \frac{g^2 \langle \bar{q}q \rangle^2}{2^{9}3^4 5\pi^8} \right) s^4 \\
			& + \left( \frac{\langle \bar{s}s \rangle^2}{2^{10}3^1 5\pi^6} + \frac{\langle \bar{q}q \rangle^2}{2^{8}3^1 5\pi^6} + \frac{\langle g^3 f G^3 \rangle}{2^{15}3^2 5 \pi^{10}} + \frac{m_s \langle g \bar{s}\sigma G s \rangle}{2^{12}3^1 5 \pi^8} \right) s^4 + \left( \frac{5 m_s \langle \bar{s}s \rangle \langle g^2 G^2 \rangle}{2^{12}3^2 \pi^8} - \frac{\langle \bar{q} q \rangle \langle g \bar{q} \sigma G q \rangle}{2^8 3 \pi^6} \right) s^3.
		\end{aligned}
		\label{eq:OPE for SigmaSigma}
	\end{equation}
\end{widetext}
This OPE expression contains almost complete contributions up to dimension 6. For the dimension-8 terms, we have performed a simple calculation under the factorization assumption. The relevant parameters are listed in Eq.~\eqref{eq:value of condensate}.

\begin{figure*}[htbp]
	\centering
	\includegraphics[width=0.5\textwidth]{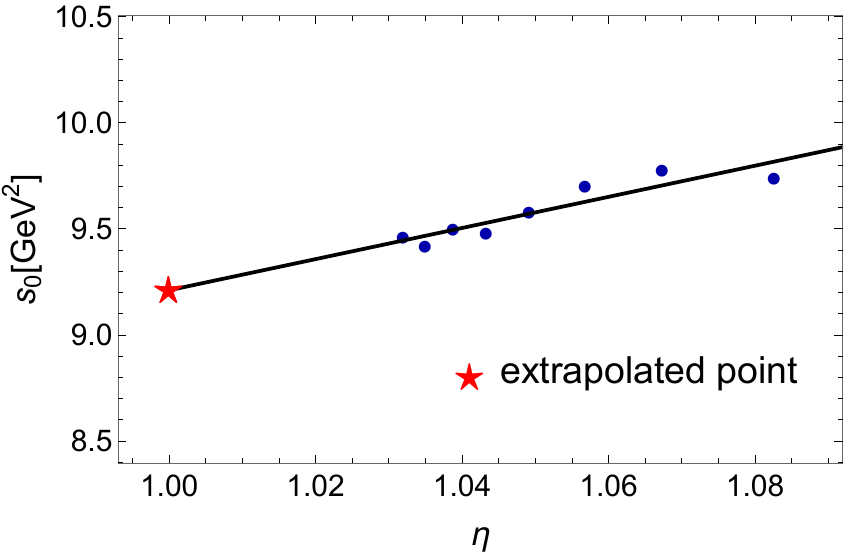}
	\caption{Linear extrapolation of the extracted minima $(s_0^\star,\eta_{\text{min}})$ to the ideal limit $\eta_{\text{min}}=1$ for $\Sigma \bar{\Sigma}$, with $\Delta n=10$, $\Delta s_0=0.02 \ \mathrm{GeV}^2$, and $\epsilon=10^{-3}$, where the $\langle g^3 f G^3\rangle$ is taken as $1.2\langle \alpha_s G^2 \rangle \ \mathrm{GeV}^2$.}
	\label{fig:searchforX(2370)}
\end{figure*}

Within the IMSR framework, we perform the numerical analysis with $n\in [40,140]$ and step size $\Delta n=10$, $\Delta s_0=0.02\ \mathrm{GeV}^2$, and $\epsilon=10^{-3}$. The extrapolation result is presented in Fig.~\ref{fig:searchforX(2370)}. The duality parameter, hadron mass, and squared coupling are finally found to be
\begin{equation}
	\label{eq:mass of high}
	\begin{aligned}
		s_0(\eta=1) & =9.21 \pm 0.03 \ \mathrm{GeV}^2, \\
		m_{\Sigma \bar{\Sigma}}&=2.74 \pm 0.12 \ \mathrm{GeV}, \quad  
		\\
		f^2_{\Sigma \bar{\Sigma}}&=(5.83\pm1.86)\times10^{-7} \ \mathrm{GeV}^{16}. 
	\end{aligned}
\end{equation}
The error in $s_0$ comes only from the fit and extrapolation, whereas those in $m_{\Sigma \bar{\Sigma}}$ and $f^2_{\Sigma \bar{\Sigma}}$ also include the uncertainties of the condensates and the strange quark mass. 

The extracted mass in Eq.~\eqref{eq:mass of high} is much higher than the mass of $X(2370)$. It is known that the trigluon condensate $\langle g^3 f G^3 \rangle$ has not been well determined to date and can only be estimated from the dimension-4 gluon condensate. In Ref.~\cite{Narison:2011rn,Narison:2011xe}, a quite different estimation was provided as
\begin{equation}
	\langle g^3 f G^3 \rangle =( 8.2 \pm 1.0 ) \,\text{GeV}^2 \times \langle \alpha_s G^2\rangle , 
\end{equation}
which is much larger than that in Eq.~\eqref{eq:value of condensate}. 
Adopting such a value of the trigluon condensate, our analysis shows that the extracted hadron mass decreases to around $2.65\ \mathrm{GeV}$, which is still much higher than the mass of $X(2370)$. These results do not support the interpretation of $X(2370)$ as a $\Sigma\bar{\Sigma}$ molecular state.

Interestingly, a very recent BESIII report on \(\chi_{cJ} \to \Sigma^0 \bar{\Sigma}^0 \eta\) shows a faint accumulation of events around \(2.7 \ \mathrm{GeV}\) in the \(\Sigma^0 \bar{\Sigma}^0\) invariant mass distributions, although the statistical significance is currently low~\cite{BESIII:2026emk}. This experimental hint roughly agrees with our results, awaiting further confirmation with higher statistics.

\begin{figure*}[htbp]
	\centering
	\includegraphics[width=0.5\textwidth]{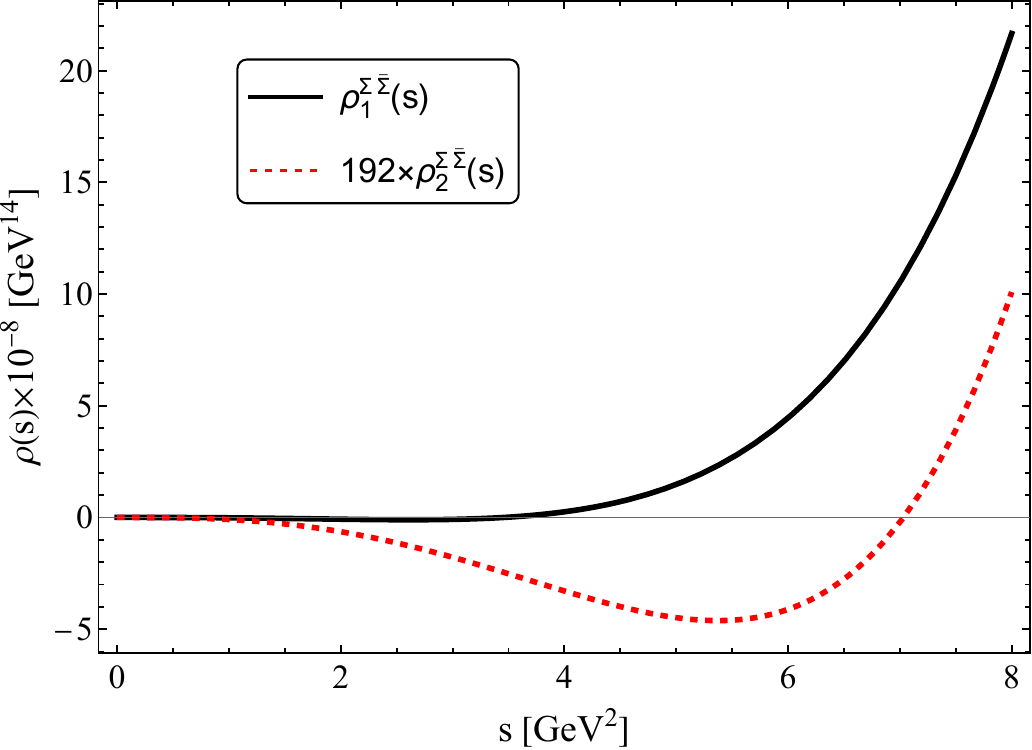}
	\caption{The spectral densities $\rho_1^{\Sigma\bar{\Sigma}}(s)$ and $192\times\rho_2^{\Sigma\bar{\Sigma}}(s)$. The factor $192$ is introduced to match their perturbative coefficients.}
	\label{fig:compare with old}
\end{figure*}

To cross-check the molecular interpretation, we investigate another current that can couple to the $\Sigma$ particle~\cite{Chung:1981wm}
\begin{equation}
	j_{\Sigma}=\epsilon_{abc}(u_a^T\mathcal{C}\gamma_5 s_b)d_c.
\end{equation}
The induced $\bar{\Sigma}\Sigma$ interpolating current is~\cite{Wan:2021vny}
\begin{equation}
  j_{\bar{\Sigma}\Sigma}=j_{\bar{\Sigma}}\gamma_5 j_{\Sigma}. \label{sigmasigma2}
\end{equation}
Up to dimension-8, the OPE-side spectral density reads
\begin{widetext}
	\begin{equation}
		\begin{aligned}
			\rho_2^{\Sigma \bar{\Sigma}}(s) = & \frac{s^7}{2^{23}3^2 5^2 7^2 \pi^{10}} 
			+ \left( \frac{\langle g^2 G^2 \rangle}{2^{22} 3^2 5^2 \pi^{10}} + \frac{m_s \langle \bar{s} s \rangle}{2^{17}3^2 5^2 \pi^8} - \frac{m_s \langle \bar{q} q \rangle}{2^{16}3^2 5^2 \pi^8} \right) s^5 \\
			& + \left(\frac{\langle \bar{q}q \rangle \langle \bar{s}s \rangle}{2^{13}3^2 5\pi^6} + \frac{\langle \bar{q}q \rangle^2}{2^{14}3^2 5\pi^6} + \frac{\langle g^3 f G^3 \rangle}{2^{21}3^3 5 \pi^{10}} + \frac{m_s \langle g \bar{s}\sigma G s \rangle}{2^{18}3^2 5 \pi^{8}} +\frac{m_s \langle g \bar{q}\sigma G q \rangle}{2^{17}3^1 5 \pi^{8}} \right) s^4
			+ \left(\frac{g^2 \langle \bar{s}s \rangle^2}{2^{15}3^5 5\pi^8} -\frac{g^2 \langle \bar{q}q \rangle^2}{2^{15}3^5 5\pi^8} \right) s^4 \\
			& + \left(\frac{m_s \langle \bar{s}s \rangle \langle g^2 G^2 \rangle}{2^{18}3^2 \pi^8} -\frac{5 m_s \langle \bar{q}q \rangle \langle g^2 G^2 \rangle}{2^{17}3^3 \pi^8} \right) s^3 
			- \left( \frac{\langle \bar{q}q\rangle \langle g\bar{q}\sigma G q\rangle + \langle \bar{s}s\rangle \langle g\bar{q}\sigma G q\rangle +\langle \bar{q}q\rangle \langle g\bar{s}\sigma G s \rangle}{2^{13} 3^2\pi^6}  \right)s^3.
		\end{aligned}
		\label{eq:OPE for SigmaSigmm but j2}
	\end{equation}
\end{widetext}
However, the sum rule analysis based on this current is not reasonable due to the pathological behavior of its spectral function. This spectral function becomes negative in the region $s<7\ \mathrm{GeV}^2$, as clearly illustrated in Fig.~\ref{fig:compare with old}. Such a bad behavior strongly violates the physical positivity of the spectral function, and thus does not support the existence of the $\bar{\Sigma}\Sigma$ state. In contrast, the spectral function in Eq.~\eqref{eq:OPE for SigmaSigma} exhibits a much better positivity behavior to provide reliable mass prediction. 

In Ref.~\cite{Wan:2021vny}, the current $j_{\bar{\Sigma}\Sigma}$ in Eq.~\eqref{sigmasigma2} has been investigated to predict the mass of the $\bar{\Sigma}\Sigma$ state to be around $2.39\ \mathrm{GeV}$, which is  consistent with the mass of $X(2370)$. The discrepancy with our result may stem from their inclusion of a subset of high-dimensional condensates up to dimension-14 in the OPE. However, it is difficult to reliably estimate the contributions of these high-dimensional condensates if only a fraction of them are calculated. Actually, the number of Feynman diagrams grows factorially as the dimension increases~\cite{Shifman:2000jv}. Consequently, incorporating such high-dimensional terms only increases theoretical uncertainty and undermines the reliability of the extracted mass.

\section{Comparison with Laplace sum rules}\label{sec:discussion}
In both the LSR and the IMSR, the contributions from higher excited states are suppressed. In the LSR, this is accomplished through the Borel transform, which introduces an exponential suppression factor. In the IMSR, the suppression arises from the $n-$th derivative, which provides a polynomial damping. As demonstrated in Fig.~\ref{fig:rho mass in Laplace} and~ Fig.~\ref{fig:rho mass in s0 moment}, the two methods yield numerically consistent results for the ground-state mass.

\begin{figure*}[htbp]
	\centering
	\includegraphics[width=0.5\textwidth]{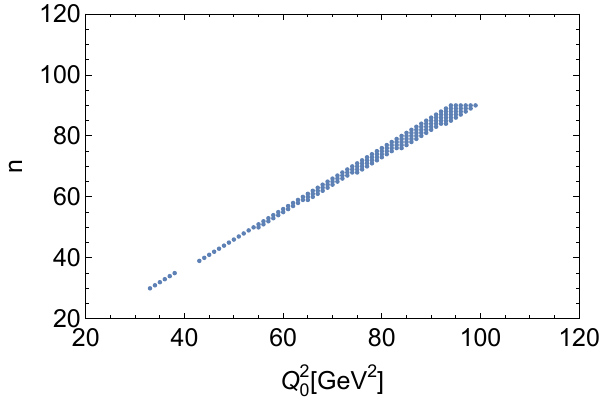}
	\caption{Dependence check for the OPE in Eq.~\eqref{eq:OPE of P6} with $\epsilon=0.25\times 10^{-3}$ and $s_0=3.42 \ \mathrm{GeV}^2$.}
	\label{fig:dependence in final}
\end{figure*}

Although both the LSR and the IMSR are founded on the principle of quark-hadron duality, a notable distinction arises in how they treat the spectral density in the high-energy region. In the LSR, the phenomenological spectral density $\rho^h(s)$ above the threshold $s_h$ is explicitly canceled by the OPE spectral density $\rho^{\text{OPE}}(s)$ above $s_0$, effectively isolating the ground-state contribution. However, this cancellation does not come with a built-in criterion for fixing the parameters; one must introduce subjective conditions to constrain the Borel window and duality parameter. 
Applying the same duality relation directly in the IMSR would lead to the same issue (Route 1 in Section~\ref{sec:two logical routes}). To overcome this, we adopt the approximation $\delta_n(s_h,Q^2_0)\simeq\delta_{n+1}(s_h,Q^2_0)$ from the conventional moment approach. Consistency with the quark-hadron duality then implies that an analogous relation $\delta'_n(s_0,Q^2_0)\simeq\delta'_{n+1}(s_0,Q^2_0)$ must also hold on the OPE side.

It is worth noting that the results for $K(1690)$ in Eqs.~\eqref{Ks0}-\eqref{Kmass} are consistent with those obtained in Ref.~\cite{Zhang:2025fuz}. However, the determination in the LSR is subject to the conventional criteria of OPE convergence and pole dominance, which are inherently subjective and introduce significant uncertainties. In contrast, the present IMSR result is obtained without invoking such ad hoc criteria, rendering it more robust and objective. Despite the subjective nature of the conventional criteria, the two methods can still yield essentially consistent results when certain conditions are met, e.g., by quantifying the LSR stability via a $\chi^2(s_0)$ function to select the duality parameter~\cite{Xu:2025oqn,Li:2025dkw,Li:2025hsp,Chen:2024bpz}.

A further perspective on the equivalence between the two methods can be gained by examining the parameter space $(Q^2_0,n)$ that satisfies the dependence condition in Eq.~\eqref{eq:dependence condition}. Specifically, we fix $s_0=3.42 \ \mathrm{GeV}^2$ and $\epsilon=0.25\times 10^{-3}$, and identify the parameter space for which the OPE expression in Eq.~\eqref{eq:OPE of P6} satisfies the dependence condition. The resulting allowed region is shown in Fig.~\ref{fig:dependence in final}, within which the ratio $Q^2_0/n$ remains nearly constant at approximately $1.1 \ \mathrm{GeV}^2$. This observation provides a direct connection between the IMSR and LSR frameworks: the constant ratio corresponds to the Borel window $1.14 \ \mathrm{GeV}^2 \leq M^2_B \leq 1.35 \ \mathrm{GeV}^2$ determined in our previous LSR analysis~\cite{Zhang:2025fuz}.

\section{Summary}\label{sec:summary}

In this work, we have developed an improved formulation of the moment sum rules. We begin by reexamining the derivations of both the LSR and conventional MSR, and identify their limitations: the conventional MSR cannot determine the coupling of the interpolating current, whereas the LSR depends on the ad hoc criteria of OPE convergence and pole dominance, which introduce subjectivity and non-uniqueness. More importantly, the two methods yield inconsistent mass predictions for the ground-state resonance. 

We demonstrate that this inconsistency originates from the absence of quark-hadron duality in the conventional MSR, which permits contamination from higher excited states on the OPE side and consequently overestimates the ground-state mass. By properly incorporating duality, we introduce an improved moment sum rule method that addresses these problems.

In this formulation, the approximation $\delta_n(s_h,Q^2_0) \simeq \delta_{n+1}(s_h,Q^2_0)$ serves as the sufficient condition for ensuring that only the ground-state resonance is retained on the phenomenological side. On the OPE side, we introduce the corresponding quantity $\delta'_n(s_0,Q^2_0)$. Consistency with quark-hadron duality then naturally requires $\delta'_n(s_0,Q^2_0) \simeq \delta'_{n+1}(s_0,Q^2_0)$, which serves as an a posteriori constraint on the parameters.
Correspondingly, the ``narrow resonance'' assumption on the phenomenological side implies model self-consistency conditions on the OPE side, thereby defining the dependence conditions. To control the deviation from the single-pole limit, we introduce a tolerance parameter $\epsilon$ in Eq.~\eqref{eq:dependence condition}.

The combination of the approximation and the dependence conditions leads to a well-defined numerical procedure that uniquely determines both the duality parameter and the ground-state mass, without invoking any ad hoc criteria. The extracted $(s_0, m_X)$ are thus essentially unique. This procedure, in which the physical outputs are fixed by self-consistency conditions, is conceptually analogous to the Bootstrap philosophy. 

We apply the IMSR method to investigate whether $K(1690)$ and $X(2370)$ can be interpreted as the pseudoscalar $ud\bar{d}\bar{s}$ tetraquark and $\Sigma\bar{\Sigma}$ molecular state, respectively. The results for the $ud\bar{d}\bar{s}$ state are consistent with those obtained from our previous LSR analysis, supporting the $K(1690)$ to be a tetraquark state and also providing strong evidence for the equivalence of the two approaches.
For the $\Sigma\bar{\Sigma}$ molecular state with $I(J^{PC})=0(0^{-+})$, the extracted mass is approximately $2.74\pm0.12\ \mathrm{GeV}$, which is inconsistent with the mass of $X(2370)$ and thus does not support its $\Sigma\bar{\Sigma}$ molecular interpretation.

These applications show that the IMSR method presented here effectively eliminates the problem of subjectivity that has long plagued the LSR, thereby significantly enhancing the robustness and reliability of QCD sum rule predictions.

\textit{\textbf{Acknowledgement.}} \textemdash \,

J.-P. Z. and X.-L. C. thank Prof.~Wenliang Li for valuable discussions. J.-P. Z. thanks Zhijian Huang, Zi-Xi Ou-Yang, Li-jing Zhang, and Wen-yun Wu for useful discussions. This work is supported by the National Natural Science Foundation of China under Grant No. 12575153.

\appendix

\section{Proof and interpretation of $\eta(s_0,Q^2_0,n) > 1$}\label{sec:eta}
In this appendix, we provide a proof to demonstrate that the approximate function $\eta(s_0,Q^2_0,n) > 1$.  Substituting the definition of $\eta(s_0,Q^2_0,n)$ into Eq.~\eqref{eq:mass of improve moment}, one obtains the result 
\begin{equation}
	\begin{aligned}
		\eta(s_0,Q^2_0,n)=\frac{1}{m^2_X+Q^2_0}\frac{\int_{s_0}^{\infty} \frac{\rho^{\text{OPE}}(s)}{(s+Q^2_0)^{n+1}}\mathrm{d}s}{\int_{s_0}^{\infty} \frac{\rho^{\text{OPE}}(s)}{(s+Q^2_0)^{n+2}}\mathrm{d}s}.
	\end{aligned}
\end{equation}
Subject to the constraint Eq.~\eqref{eq:dependence condition}, $m^2_X$ can be treated as approximately independent of \(n\). In the limit of $n \rightarrow \infty$, the function $\eta$ approaches unity for any physical spectral function $\rho^{\text{OPE}}(s)$. If we can establish that $\eta$ is monotonically decreasing with respect to $n$, it follows directly that $\eta>1$ for any finite $n>0$.
We investigate the derivative $\mathrm{d}\eta / \mathrm{d} n$ to determine its monotonicity (omitting a positive factor)
\begin{equation}
	\begin{aligned}
		\frac{\mathrm{d}\eta}{\mathrm{d} n} \propto  \int_{s_0}^{\infty} \frac{\rho^{\text{OPE}}(s) \mathrm{d}s}{(s+Q^2_0)^{n+1}} \int_{s_0}^{\infty} \frac{\rho^{\text{OPE}}(s) \ln(s+Q^2_0) \mathrm{d}s}{(s+Q^2_0)^{n+2}}
		\\  -
		\int_{s_0}^{\infty} \frac{\rho^{\text{OPE}}(s) \mathrm{d}s}{(s+Q^2_0)^{n+2}} \int_{s_0}^{\infty} \frac{\rho^{\text{OPE}}(s) \ln(s+Q^2_0) \mathrm{d}s}{(s+Q^2_0)^{n+1}}.
	\end{aligned} \label{eq:derivative eta/n}
\end{equation}
We introduce the variable substitution $x=s+Q^2_0$, which changes the integration lower limit to $x_0=s_0+Q^2_0$. In terms of this new variable, Eq.~\eqref{eq:derivative eta/n} can be expressed compactly by defining the positive weight function $W(x)$ as
\begin{equation}
	\begin{aligned}
		W(x)=\frac{\rho^{\text{OPE}}(x-Q^2_0)}{x^{n+2}}.
	\end{aligned} 
\end{equation}
Then one can express Eq.~\eqref{eq:derivative eta/n} as
\begin{equation}
	\begin{aligned}
		\frac{\mathrm{d}\eta}{\mathrm{d} n} \propto  \int_{x_0}^\infty W(x) \, x \, \mathrm{d} x\int_{x_0}^\infty W(x) \, \ln x \, \mathrm{d}x \\ -
		\int_{x_0}^\infty W(x) \, \mathrm{d} x \int_{x_0}^\infty W(x) \, x \ln x \, \mathrm{d}x.
	\end{aligned} 
\end{equation}

We then apply Chebyshev's integral inequality to determine the sign of $\mathrm{d}\eta / \mathrm{d} n$. For two functions $f(x)$ and $g(x)$ of identical monotonicity on the interval $[a,b]$, and a weight function $W(x)\ge 0$, Chebyshev's integral inequality takes the form
\begin{equation}
	\begin{aligned} \label{Chebyshev inequality}
		\int_{a}^{b} W(x) \mathrm{d} x \int_{a}^{b} W(x)f(x)g(x)\mathrm{d}x \\ \ge \int_{a}^{b} W(x) f(x) \mathrm{d} x \int_{a}^{b} W(x) g(x)\mathrm{d}x,
	\end{aligned} 
\end{equation}
in which the equality holds only if at least one of the functions $f(x)$ and $g(x)$ is constant.

Before applying Chebyshev's inequality, we verify its prerequisites: $Q^2_0>0$, $n>0$, and $n$ is chosen larger than the highest power in $\rho^{\text{OPE}}(s)$ to ensure convergence. For $s>s_0$, $\rho^{\text{OPE}}(s)>0$, which ensures that the weight function $W(x)$ is positive-definite. This is physically guaranteed; otherwise, the ground-state mass would be determined by an unphysical negative-definite spectral density.

To apply Chebyshev's integral inequality in Eq.~\eqref{Chebyshev inequality}, we define $f(x)=x$ and $g(x)=\ln(x)$, both of which are strictly increasing on the interval $[x_0,\infty)$, and hence they share the same monotonicity. The conditions required by the inequality are therefore fully satisfied, and we obtain
\begin{equation}
	\begin{aligned} 
		\int_{x_0}^{\infty} W(x) \mathrm{d} x \int_{x_0}^{\infty} W(x)x \ln x \mathrm{d}x  \\ > \int_{x_0}^{\infty} W(x) x \mathrm{d} x \int_{x_0}^{\infty} W(x) \ln x \mathrm{d}x.
	\end{aligned} \label{eq:Chebyshev in moment}
\end{equation}

According to Eq.~\eqref{eq:derivative eta/n}, this result leads to 
\begin{equation}
	\begin{aligned} 
		\frac{\mathrm{d} \eta}{\mathrm{d} n} < 0,
	\end{aligned} 
\end{equation}
implying that $\eta(s_0,Q^2_0,n)$ is strictly monotonically decreasing with respect to $n$, while $\eta>1$ for any finite $n>0$ (provided integral convergence is at least ensured).

\section{Statistical meaning of the dependence conditions}
\label{sec:dependence meaning}
In this appendix we provide the detailed derivations of the two relations in Eq.~\eqref{eq:differentiating with Q and n}. With the definition \(x=s+Q_0^2\), Eq.~\eqref{eq:mass equation with tjx meaning} becomes
\begin{equation}
	m^2_X+Q^2_0=E_n^{-1}[x^{-1}]. \label{eq:mass equation with tjx meaning 2}
\end{equation}
Since \(\partial/\partial Q_0^2 = \partial/\partial x\), we compute the following derivative using the quotient rule
\begin{equation}
	\begin{aligned} 
	\frac{\partial}{\partial x}
	E_n[x^{-1}]
	&=
	\frac{\partial}{\partial x}
	\frac{\int \rho(s)\,x^{-n-2}\,\mathrm{d}s}
	{\int \rho(s)\,x^{-n-1}\,\mathrm{d}s} \\
	&=
	-(n+2)\,\mathrm{Var}_n[x^{-1}]
	-
	E_n^2[x^{-1}],
	\end{aligned} 
\end{equation}
where \(\mathrm{Var}_n\) denotes the variance under the \(\mathrm{d}\mu_n\) in Eq.~\eqref{eq:weighted function} 
\begin{equation}
	\mathrm{Var}_n[x^{-1}]
	=
	E_n[x^{-2}]
	-
	E_n^2[x^{-1}]. \label{variance}
\end{equation}
Therefore,
\begin{align}
	\frac{\partial}{\partial Q_0^2}&
	E_n^{-1}[x^{-1}]
	=
	-
	E_n^{-2}[x^{-1}]
	\frac{\partial}{\partial Q_0^2}
	E_n[x^{-1}]  \nonumber \\
	&=
	E_n^{-2}[x^{-1}]
	\left[
	(n+2)\,\mathrm{Var}_n[x^{-1}]
	+
	E_n^2[x^{-1}]
	\right] .
\end{align}
On the other hand
\begin{equation}
	\frac{\partial}{\partial Q_0^2}
	\left(m_X^2+Q_0^2\right)
	=
	\frac{\partial m_X^2}{\partial Q_0^2}
	+
	1 .
\end{equation}
Equating the two expressions and using Eq.~\eqref{eq:mass equation with tjx meaning 2}, we arrive at
\begin{equation}
		\frac{\partial m_X^2}{\partial Q_0^2}
		=
		(n+2)(m_X^2+Q_0^2)^2\,
		\mathrm{Var}_n\!\left[\frac{1}{s+Q_0^2}\right]
	\label{eq:derivative Q app}
\end{equation}

Now we compute \(\partial m_X^2/\partial n\). We first calculate
\begin{equation}
	\frac{\partial}{\partial n}
	E_n[x^{-1}]
	=
	\frac{\partial}{\partial n}
	\left(
	\frac{\tilde{M}_{n+1}}{\tilde{M}_n}
	\right) .
\end{equation}
The derivative of \(\tilde{M}_n\) with respect to \(n\) is
\begin{equation}
	\frac{\partial \tilde{M}_n}{\partial n}
	=
	-
	\tilde{M}_n\,
	E_n[\ln x] .
\end{equation}
Then
\begin{align}
	\frac{\partial}{\partial n}
	E_n[x^{-1}]
	&=
	\frac{
		\left(\partial_n \tilde{M}_{n+1}\right)\tilde{M}_n
		-
		\tilde{M}_{n+1}\left(\partial_n \tilde{M}_n\right)
	}{
		\tilde{M}_n^2
	} \nonumber \\[4pt]
	&=
	\frac{\tilde{M}_{n+1}}{\tilde{M}_n}
	\left(
	E_n[\ln x]
	-
	E_{n+1}[\ln x]
	\right) .
\end{align}
Using the relation
\begin{equation}
	E_{n+1}[f]
	= E_n\left[\frac{f}{x}\right]E^{-1}_n[x^{-1}],
\end{equation}
we have
\begin{equation}
	E_{n+1}[\ln x]
	= E_n\left[\frac{\ln x}{x}\right]E^{-1}_n[x^{-1}].
\end{equation}
Therefore
\begin{equation}
	\frac{\partial}{\partial n}
	E_n[x^{-1}]
	=
	E_n[x^{-1}]
	E_n[\ln x]
	-
	E_n\left[\frac{\ln x}{x}\right]
	=
	-
	\mathrm{Cov}_n\left[\frac{1}{x},\ln x\right] ,
\end{equation}
where we have used the definition of the covariance
\begin{equation}
	\mathrm{Cov}[X,Y]=E[XY]-E[X]E[Y]. \label{covariance}
\end{equation}
Thus
\begin{equation}
	\begin{aligned}
		& \frac{\partial}{\partial n}  \left (m_X^2+Q_0^2\right)
		= \frac{\partial m_X^2}{\partial n} 
		= \frac{\partial}{\partial n} E_n^{-1}[x^{-1}] \\
		&= -E_n^{-2}[x^{-1}] \frac{\partial}{\partial n} E_n[x^{-1}] = E_n^{-2}[x^{-1}] \mathrm{Cov}_n\left[\frac{1}{x},\ln x\right] .
	\end{aligned}
\end{equation}
Finally, using Eq.~\eqref{eq:mass equation with tjx meaning 2} and the definition of $x$, we obtain
\begin{equation}
		\frac{\partial m_X^2}{\partial n}
		=
		(m_X^2+Q_0^2)^2\,
		\mathrm{Cov}_n\!\left[\frac{1}{s+Q_0^2},\ln(s+Q_0^2)\right]
\end{equation}
Therefore, the dependence conditions are of purely statistical origin, determined by the variance and covariance of the weighted spectral function in Eq.~\eqref{eq:weighted function}.


\begin{thebibliography}{57}%
	\makeatletter
	\providecommand \@ifxundefined [1]{%
		\@ifx{#1\undefined}
	}%
	\providecommand \@ifnum [1]{%
		\ifnum #1\expandafter \@firstoftwo
		\else \expandafter \@secondoftwo
		\fi
	}%
	\providecommand \@ifx [1]{%
		\ifx #1\expandafter \@firstoftwo
		\else \expandafter \@secondoftwo
		\fi
	}%
	\providecommand \natexlab [1]{#1}%
	\providecommand \enquote  [1]{``#1''}%
	\providecommand \bibnamefont  [1]{#1}%
	\providecommand \bibfnamefont [1]{#1}%
	\providecommand \citenamefont [1]{#1}%
	\providecommand \href@noop [0]{\@secondoftwo}%
	\providecommand \href [0]{\begingroup \@sanitize@url \@href}%
	\providecommand \@href[1]{\@@startlink{#1}\@@href}%
	\providecommand \@@href[1]{\endgroup#1\@@endlink}%
	\providecommand \@sanitize@url [0]{\catcode `\\12\catcode `\$12\catcode
		`\&12\catcode `\#12\catcode `\^12\catcode `\_12\catcode `\%12\relax}%
	\providecommand \@@startlink[1]{}%
	\providecommand \@@endlink[0]{}%
	\providecommand \url  [0]{\begingroup\@sanitize@url \@url }%
	\providecommand \@url [1]{\endgroup\@href {#1}{\urlprefix }}%
	\providecommand \urlprefix  [0]{URL }%
	\providecommand \Eprint [0]{\href }%
	\providecommand \doibase [0]{https://doi.org/}%
	\providecommand \selectlanguage [0]{\@gobble}%
	\providecommand \bibinfo  [0]{\@secondoftwo}%
	\providecommand \bibfield  [0]{\@secondoftwo}%
	\providecommand \translation [1]{[#1]}%
	\providecommand \BibitemOpen [0]{}%
	\providecommand \bibitemStop [0]{}%
	\providecommand \bibitemNoStop [0]{.\EOS\space}%
	\providecommand \EOS [0]{\spacefactor3000\relax}%
	\providecommand \BibitemShut  [1]{\csname bibitem#1\endcsname}%
	\let\auto@bib@innerbib\@empty
	\bibitem [{\citenamefont {Shifman}\ \emph
		{et~al.}(1979{\natexlab{a}})\citenamefont {Shifman}, \citenamefont
		{Vainshtein},\ and\ \citenamefont {Zakharov}}]{Shifman:1978bx}%
	\BibitemOpen
	\bibfield  {author} {\bibinfo {author} {\bibfnamefont {M.~A.}\ \bibnamefont
			{Shifman}}, \bibinfo {author} {\bibfnamefont {A.~I.}\ \bibnamefont
			{Vainshtein}},\ and\ \bibinfo {author} {\bibfnamefont {V.~I.}\ \bibnamefont
			{Zakharov}},\ }\bibfield  {title} {\bibinfo {title} {{QCD and Resonance
				Physics. Theoretical Foundations}},\ }\href
	{https://doi.org/10.1016/0550-3213(79)90022-1} {\bibfield  {journal}
		{\bibinfo  {journal} {Nucl. Phys. B}\ }\textbf {\bibinfo {volume} {147}},\
		\bibinfo {pages} {385} (\bibinfo {year} {1979}{\natexlab{a}})}\BibitemShut
	{NoStop}%
	\bibitem [{\citenamefont {Shifman}\ \emph
		{et~al.}(1979{\natexlab{b}})\citenamefont {Shifman}, \citenamefont
		{Vainshtein},\ and\ \citenamefont {Zakharov}}]{Shifman:1978by}%
	\BibitemOpen
	\bibfield  {author} {\bibinfo {author} {\bibfnamefont {M.~A.}\ \bibnamefont
			{Shifman}}, \bibinfo {author} {\bibfnamefont {A.~I.}\ \bibnamefont
			{Vainshtein}},\ and\ \bibinfo {author} {\bibfnamefont {V.~I.}\ \bibnamefont
			{Zakharov}},\ }\bibfield  {title} {\bibinfo {title} {{QCD and Resonance
				Physics: Applications}},\ }\href
	{https://doi.org/10.1016/0550-3213(79)90023-3} {\bibfield  {journal}
		{\bibinfo  {journal} {Nucl. Phys. B}\ }\textbf {\bibinfo {volume} {147}},\
		\bibinfo {pages} {448} (\bibinfo {year} {1979}{\natexlab{b}})}\BibitemShut
	{NoStop}%
	\bibitem [{\citenamefont {Reinders}\ \emph {et~al.}(1985)\citenamefont
		{Reinders}, \citenamefont {Rubinstein},\ and\ \citenamefont
		{Yazaki}}]{Reinders:1984sr}%
	\BibitemOpen
	\bibfield  {author} {\bibinfo {author} {\bibfnamefont {L.~J.}\ \bibnamefont
			{Reinders}}, \bibinfo {author} {\bibfnamefont {H.}~\bibnamefont
			{Rubinstein}},\ and\ \bibinfo {author} {\bibfnamefont {S.}~\bibnamefont
			{Yazaki}},\ }\bibfield  {title} {\bibinfo {title} {{Hadron Properties from
				QCD Sum Rules}},\ }\href {https://doi.org/10.1016/0370-1573(85)90065-1}
	{\bibfield  {journal} {\bibinfo  {journal} {Phys. Rept.}\ }\textbf {\bibinfo
			{volume} {127}},\ \bibinfo {pages} {1} (\bibinfo {year} {1985})}\BibitemShut
	{NoStop}%
	\bibitem [{\citenamefont {Chen}\ \emph {et~al.}(2016)\citenamefont {Chen},
		\citenamefont {Chen}, \citenamefont {Liu},\ and\ \citenamefont
		{Zhu}}]{Chen:2016qju}%
	\BibitemOpen
	\bibfield  {author} {\bibinfo {author} {\bibfnamefont {H.-X.}\ \bibnamefont
			{Chen}}, \bibinfo {author} {\bibfnamefont {W.}~\bibnamefont {Chen}}, \bibinfo
		{author} {\bibfnamefont {X.}~\bibnamefont {Liu}},\ and\ \bibinfo {author}
		{\bibfnamefont {S.-L.}\ \bibnamefont {Zhu}},\ }\bibfield  {title} {\bibinfo
		{title} {{The hidden-charm pentaquark and tetraquark states}},\ }\href
	{https://doi.org/10.1016/j.physrep.2016.05.004} {\bibfield  {journal}
		{\bibinfo  {journal} {Phys. Rept.}\ }\textbf {\bibinfo {volume} {639}},\
		\bibinfo {pages} {1} (\bibinfo {year} {2016})},\ \Eprint
	{https://arxiv.org/abs/1601.02092} {arXiv:1601.02092 [hep-ph]} \BibitemShut
	{NoStop}%
	\bibitem [{\citenamefont {Lebed}\ \emph {et~al.}(2017)\citenamefont {Lebed},
		\citenamefont {Mitchell},\ and\ \citenamefont {Swanson}}]{Lebed:2016hpi}%
	\BibitemOpen
	\bibfield  {author} {\bibinfo {author} {\bibfnamefont {R.~F.}\ \bibnamefont
			{Lebed}}, \bibinfo {author} {\bibfnamefont {R.~E.}\ \bibnamefont
			{Mitchell}},\ and\ \bibinfo {author} {\bibfnamefont {E.~S.}\ \bibnamefont
			{Swanson}},\ }\bibfield  {title} {\bibinfo {title} {{Heavy-Quark QCD
				Exotica}},\ }\href {https://doi.org/10.1016/j.ppnp.2016.11.003} {\bibfield
		{journal} {\bibinfo  {journal} {Prog. Part. Nucl. Phys.}\ }\textbf {\bibinfo
			{volume} {93}},\ \bibinfo {pages} {143} (\bibinfo {year} {2017})},\ \Eprint
	{https://arxiv.org/abs/1610.04528} {arXiv:1610.04528 [hep-ph]} \BibitemShut
	{NoStop}%
	\bibitem [{\citenamefont {Esposito}\ \emph {et~al.}(2017)\citenamefont
		{Esposito}, \citenamefont {Pilloni},\ and\ \citenamefont
		{Polosa}}]{Esposito:2016noz}%
	\BibitemOpen
	\bibfield  {author} {\bibinfo {author} {\bibfnamefont {A.}~\bibnamefont
			{Esposito}}, \bibinfo {author} {\bibfnamefont {A.}~\bibnamefont {Pilloni}},\
		and\ \bibinfo {author} {\bibfnamefont {A.~D.}\ \bibnamefont {Polosa}},\
	}\bibfield  {title} {\bibinfo {title} {{Multiquark Resonances}},\ }\href
	{https://doi.org/10.1016/j.physrep.2016.11.002} {\bibfield  {journal}
		{\bibinfo  {journal} {Phys. Rept.}\ }\textbf {\bibinfo {volume} {668}},\
		\bibinfo {pages} {1} (\bibinfo {year} {2017})},\ \Eprint
	{https://arxiv.org/abs/1611.07920} {arXiv:1611.07920 [hep-ph]} \BibitemShut
	{NoStop}%
	\bibitem [{\citenamefont {Guo}\ \emph {et~al.}(2018)\citenamefont {Guo},
		\citenamefont {Hanhart}, \citenamefont {Mei{\ss}ner}, \citenamefont {Wang},
		\citenamefont {Zhao},\ and\ \citenamefont {Zou}}]{Guo:2017jvc}%
	\BibitemOpen
	\bibfield  {author} {\bibinfo {author} {\bibfnamefont {F.-K.}\ \bibnamefont
			{Guo}}, \bibinfo {author} {\bibfnamefont {C.}~\bibnamefont {Hanhart}},
		\bibinfo {author} {\bibfnamefont {U.-G.}\ \bibnamefont {Mei{\ss}ner}},
		\bibinfo {author} {\bibfnamefont {Q.}~\bibnamefont {Wang}}, \bibinfo {author}
		{\bibfnamefont {Q.}~\bibnamefont {Zhao}},\ and\ \bibinfo {author}
		{\bibfnamefont {B.-S.}\ \bibnamefont {Zou}},\ }\bibfield  {title} {\bibinfo
		{title} {{Hadronic molecules}},\ }\href
	{https://doi.org/10.1103/RevModPhys.90.015004} {\bibfield  {journal}
		{\bibinfo  {journal} {Rev. Mod. Phys.}\ }\textbf {\bibinfo {volume} {90}},\
		\bibinfo {pages} {015004} (\bibinfo {year} {2018})},\ \bibinfo {note}
	{[Erratum: Rev.Mod.Phys. 94, 029901 (2022)]},\ \Eprint
	{https://arxiv.org/abs/1705.00141} {arXiv:1705.00141 [hep-ph]} \BibitemShut
	{NoStop}%
	\bibitem [{\citenamefont {Liu}\ \emph {et~al.}(2019)\citenamefont {Liu},
		\citenamefont {Chen}, \citenamefont {Chen}, \citenamefont {Liu},\ and\
		\citenamefont {Zhu}}]{Liu:2019zoy}%
	\BibitemOpen
	\bibfield  {author} {\bibinfo {author} {\bibfnamefont {Y.-R.}\ \bibnamefont
			{Liu}}, \bibinfo {author} {\bibfnamefont {H.-X.}\ \bibnamefont {Chen}},
		\bibinfo {author} {\bibfnamefont {W.}~\bibnamefont {Chen}}, \bibinfo {author}
		{\bibfnamefont {X.}~\bibnamefont {Liu}},\ and\ \bibinfo {author}
		{\bibfnamefont {S.-L.}\ \bibnamefont {Zhu}},\ }\bibfield  {title} {\bibinfo
		{title} {{Pentaquark and Tetraquark states}},\ }\href
	{https://doi.org/10.1016/j.ppnp.2019.04.003} {\bibfield  {journal} {\bibinfo
			{journal} {Prog. Part. Nucl. Phys.}\ }\textbf {\bibinfo {volume} {107}},\
		\bibinfo {pages} {237} (\bibinfo {year} {2019})},\ \Eprint
	{https://arxiv.org/abs/1903.11976} {arXiv:1903.11976 [hep-ph]} \BibitemShut
	{NoStop}%
	\bibitem [{\citenamefont {Brambilla}\ \emph {et~al.}(2020)\citenamefont
		{Brambilla}, \citenamefont {Eidelman}, \citenamefont {Hanhart}, \citenamefont
		{Nefediev}, \citenamefont {Shen}, \citenamefont {Thomas}, \citenamefont
		{Vairo},\ and\ \citenamefont {Yuan}}]{Brambilla:2019esw}%
	\BibitemOpen
	\bibfield  {author} {\bibinfo {author} {\bibfnamefont {N.}~\bibnamefont
			{Brambilla}}, \bibinfo {author} {\bibfnamefont {S.}~\bibnamefont {Eidelman}},
		\bibinfo {author} {\bibfnamefont {C.}~\bibnamefont {Hanhart}}, \bibinfo
		{author} {\bibfnamefont {A.}~\bibnamefont {Nefediev}}, \bibinfo {author}
		{\bibfnamefont {C.-P.}\ \bibnamefont {Shen}}, \bibinfo {author}
		{\bibfnamefont {C.~E.}\ \bibnamefont {Thomas}}, \bibinfo {author}
		{\bibfnamefont {A.}~\bibnamefont {Vairo}},\ and\ \bibinfo {author}
		{\bibfnamefont {C.-Z.}\ \bibnamefont {Yuan}},\ }\bibfield  {title} {\bibinfo
		{title} {{The $XYZ$ states: experimental and theoretical status and
				perspectives}},\ }\href {https://doi.org/10.1016/j.physrep.2020.05.001}
	{\bibfield  {journal} {\bibinfo  {journal} {Phys. Rept.}\ }\textbf {\bibinfo
			{volume} {873}},\ \bibinfo {pages} {1} (\bibinfo {year} {2020})},\ \Eprint
	{https://arxiv.org/abs/1907.07583} {arXiv:1907.07583 [hep-ex]} \BibitemShut
	{NoStop}%
	\bibitem [{\citenamefont {Chen}\ \emph {et~al.}(2023)\citenamefont {Chen},
		\citenamefont {Chen}, \citenamefont {Liu}, \citenamefont {Liu},\ and\
		\citenamefont {Zhu}}]{Chen:2022asf}%
	\BibitemOpen
	\bibfield  {author} {\bibinfo {author} {\bibfnamefont {H.-X.}\ \bibnamefont
			{Chen}}, \bibinfo {author} {\bibfnamefont {W.}~\bibnamefont {Chen}}, \bibinfo
		{author} {\bibfnamefont {X.}~\bibnamefont {Liu}}, \bibinfo {author}
		{\bibfnamefont {Y.-R.}\ \bibnamefont {Liu}},\ and\ \bibinfo {author}
		{\bibfnamefont {S.-L.}\ \bibnamefont {Zhu}},\ }\bibfield  {title} {\bibinfo
		{title} {{An updated review of the new hadron states}},\ }\href
	{https://doi.org/10.1088/1361-6633/aca3b6} {\bibfield  {journal} {\bibinfo
			{journal} {Rept. Prog. Phys.}\ }\textbf {\bibinfo {volume} {86}},\ \bibinfo
		{pages} {026201} (\bibinfo {year} {2023})},\ \Eprint
	{https://arxiv.org/abs/2204.02649} {arXiv:2204.02649 [hep-ph]} \BibitemShut
	{NoStop}%
	\bibitem [{\citenamefont {Meng}\ \emph {et~al.}(2023)\citenamefont {Meng},
		\citenamefont {Wang}, \citenamefont {Wang},\ and\ \citenamefont
		{Zhu}}]{Meng:2022ozq}%
	\BibitemOpen
	\bibfield  {author} {\bibinfo {author} {\bibfnamefont {L.}~\bibnamefont
			{Meng}}, \bibinfo {author} {\bibfnamefont {B.}~\bibnamefont {Wang}}, \bibinfo
		{author} {\bibfnamefont {G.-J.}\ \bibnamefont {Wang}},\ and\ \bibinfo
		{author} {\bibfnamefont {S.-L.}\ \bibnamefont {Zhu}},\ }\bibfield  {title}
	{\bibinfo {title} {{Chiral perturbation theory for heavy hadrons and chiral
				effective field theory for heavy hadronic molecules}},\ }\href
	{https://doi.org/10.1016/j.physrep.2023.04.003} {\bibfield  {journal}
		{\bibinfo  {journal} {Phys. Rept.}\ }\textbf {\bibinfo {volume} {1019}},\
		\bibinfo {pages} {1} (\bibinfo {year} {2023})},\ \Eprint
	{https://arxiv.org/abs/2204.08716} {arXiv:2204.08716 [hep-ph]} \BibitemShut
	{NoStop}%
	\bibitem [{\citenamefont {Liu}\ \emph {et~al.}(2025)\citenamefont {Liu},
		\citenamefont {Pan}, \citenamefont {Liu}, \citenamefont {Wu}, \citenamefont
		{Lu},\ and\ \citenamefont {Geng}}]{Liu:2024uxn}%
	\BibitemOpen
	\bibfield  {author} {\bibinfo {author} {\bibfnamefont {M.-Z.}\ \bibnamefont
			{Liu}}, \bibinfo {author} {\bibfnamefont {Y.-W.}\ \bibnamefont {Pan}},
		\bibinfo {author} {\bibfnamefont {Z.-W.}\ \bibnamefont {Liu}}, \bibinfo
		{author} {\bibfnamefont {T.-W.}\ \bibnamefont {Wu}}, \bibinfo {author}
		{\bibfnamefont {J.-X.}\ \bibnamefont {Lu}},\ and\ \bibinfo {author}
		{\bibfnamefont {L.-S.}\ \bibnamefont {Geng}},\ }\bibfield  {title} {\bibinfo
		{title} {{Three ways to decipher the nature of exotic hadrons: Multiplets,
				three-body hadronic molecules, and correlation functions}},\ }\href
	{https://doi.org/10.1016/j.physrep.2024.12.001} {\bibfield  {journal}
		{\bibinfo  {journal} {Phys. Rept.}\ }\textbf {\bibinfo {volume} {1108}},\
		\bibinfo {pages} {1} (\bibinfo {year} {2025})},\ \Eprint
	{https://arxiv.org/abs/2404.06399} {arXiv:2404.06399 [hep-ph]} \BibitemShut
	{NoStop}%
	\bibitem [{\citenamefont {Wang}(2026)}]{Wang:2025sic}%
	\BibitemOpen
	\bibfield  {author} {\bibinfo {author} {\bibfnamefont {Z.-G.}\ \bibnamefont
			{Wang}},\ }\bibfield  {title} {\bibinfo {title} {{Review of the QCD sum rules
				for exotic states}},\ }\href {https://doi.org/10.15302/frontphys.2026.016300}
	{\bibfield  {journal} {\bibinfo  {journal} {Front. Phys. (Beijing)}\ }\textbf
		{\bibinfo {volume} {21}},\ \bibinfo {pages} {016300} (\bibinfo {year}
		{2026})},\ \Eprint {https://arxiv.org/abs/2502.11351} {arXiv:2502.11351
		[hep-ph]} \BibitemShut {NoStop}%
	\bibitem [{\citenamefont {Leinweber}(1997)}]{Leinweber:1995fn}%
	\BibitemOpen
	\bibfield  {author} {\bibinfo {author} {\bibfnamefont {D.~B.}\ \bibnamefont
			{Leinweber}},\ }\bibfield  {title} {\bibinfo {title} {{QCD sum rules for
				skeptics}},\ }\href {https://doi.org/10.1006/aphy.1996.5641} {\bibfield
		{journal} {\bibinfo  {journal} {Annals Phys.}\ }\textbf {\bibinfo {volume}
			{254}},\ \bibinfo {pages} {328} (\bibinfo {year} {1997})},\ \Eprint
	{https://arxiv.org/abs/nucl-th/9510051} {arXiv:nucl-th/9510051} \BibitemShut
	{NoStop}%
	\bibitem [{\citenamefont {Carvunis}\ \emph {et~al.}(2024)\citenamefont
		{Carvunis}, \citenamefont {Mahmoudi},\ and\ \citenamefont
		{Monceaux}}]{Carvunis:2024koh}%
	\BibitemOpen
	\bibfield  {author} {\bibinfo {author} {\bibfnamefont {A.}~\bibnamefont
			{Carvunis}}, \bibinfo {author} {\bibfnamefont {F.}~\bibnamefont {Mahmoudi}},\
		and\ \bibinfo {author} {\bibfnamefont {Y.}~\bibnamefont {Monceaux}},\
	}\bibfield  {title} {\bibinfo {title} {{Potential of light-cone sum rules
				without semiglobal quark-hadron duality}},\ }\href
	{https://doi.org/10.1103/PhysRevD.110.114008} {\bibfield  {journal} {\bibinfo
			{journal} {Phys. Rev. D}\ }\textbf {\bibinfo {volume} {110}},\ \bibinfo
		{pages} {114008} (\bibinfo {year} {2024})},\ \Eprint
	{https://arxiv.org/abs/2404.01290} {arXiv:2404.01290 [hep-ph]} \BibitemShut
	{NoStop}%
	\bibitem [{\citenamefont {Li}\ and\ \citenamefont {Umeeda}(2020)}]{Li:2020ejs}%
	\BibitemOpen
	\bibfield  {author} {\bibinfo {author} {\bibfnamefont {H.-n.}\ \bibnamefont
			{Li}}\ and\ \bibinfo {author} {\bibfnamefont {H.}~\bibnamefont {Umeeda}},\
	}\bibfield  {title} {\bibinfo {title} {{QCD sum rules with spectral densities
				solved in inverse problems}},\ }\href
	{https://doi.org/10.1103/PhysRevD.102.114014} {\bibfield  {journal} {\bibinfo
			{journal} {Phys. Rev. D}\ }\textbf {\bibinfo {volume} {102}},\ \bibinfo
		{pages} {114014} (\bibinfo {year} {2020})},\ \Eprint
	{https://arxiv.org/abs/2006.16593} {arXiv:2006.16593 [hep-ph]} \BibitemShut
	{NoStop}%
	\bibitem [{\citenamefont {Xiong}\ \emph {et~al.}(2022)\citenamefont {Xiong},
		\citenamefont {Wei},\ and\ \citenamefont {Yu}}]{Xiong:2022uwj}%
	\BibitemOpen
	\bibfield  {author} {\bibinfo {author} {\bibfnamefont {A.-S.}\ \bibnamefont
			{Xiong}}, \bibinfo {author} {\bibfnamefont {T.}~\bibnamefont {Wei}},\ and\
		\bibinfo {author} {\bibfnamefont {F.-S.}\ \bibnamefont {Yu}},\ }\bibfield
	{title} {\bibinfo {title} {{Inverse Problem Approach for Non-Perturbative
				QCD: Foundation}},\ }\href@noop {} {\  (\bibinfo {year} {2022})},\ \Eprint
	{https://arxiv.org/abs/2211.13753} {arXiv:2211.13753 [hep-th]} \BibitemShut
	{NoStop}%
	\bibitem [{\citenamefont {Zheng}\ \emph {et~al.}(2026)\citenamefont {Zheng},
		\citenamefont {Xiong}, \citenamefont {Yu},\ and\ \citenamefont
		{Wei}}]{Zheng:2026zve}%
	\BibitemOpen
	\bibfield  {author} {\bibinfo {author} {\bibfnamefont {Y.}~\bibnamefont
			{Zheng}}, \bibinfo {author} {\bibfnamefont {A.-S.}\ \bibnamefont {Xiong}},
		\bibinfo {author} {\bibfnamefont {F.-S.}\ \bibnamefont {Yu}},\ and\ \bibinfo
		{author} {\bibfnamefont {T.}~\bibnamefont {Wei}},\ }\bibfield  {title}
	{\bibinfo {title} {{Inverse Problem Approach for Non-Perturbative QCD:
				Applications}},\ }\href@noop {} {\  (\bibinfo {year} {2026})},\ \Eprint
	{https://arxiv.org/abs/2610.40111} {arXiv:2610.40111 [hep-ph]} \BibitemShut
	{NoStop}%
	\bibitem [{\citenamefont {Li}(2021)}]{Li:2021gsx}%
	\BibitemOpen
	\bibfield  {author} {\bibinfo {author} {\bibfnamefont {H.-n.}\ \bibnamefont
			{Li}},\ }\bibfield  {title} {\bibinfo {title} {{Dispersive analysis of
				glueball masses}},\ }\href {https://doi.org/10.1103/PhysRevD.104.114017}
	{\bibfield  {journal} {\bibinfo  {journal} {Phys. Rev. D}\ }\textbf {\bibinfo
			{volume} {104}},\ \bibinfo {pages} {114017} (\bibinfo {year} {2021})},\
	\Eprint {https://arxiv.org/abs/2109.04956} {arXiv:2109.04956 [hep-ph]}
	\BibitemShut {NoStop}%
	\bibitem [{\citenamefont {Li}(2022)}]{Li:2022qul}%
	\BibitemOpen
	\bibfield  {author} {\bibinfo {author} {\bibfnamefont {H.-n.}\ \bibnamefont
			{Li}},\ }\bibfield  {title} {\bibinfo {title} {{Dispersive derivation of the
				pion distribution amplitude}},\ }\href
	{https://doi.org/10.1103/PhysRevD.106.034015} {\bibfield  {journal} {\bibinfo
			{journal} {Phys. Rev. D}\ }\textbf {\bibinfo {volume} {106}},\ \bibinfo
		{pages} {034015} (\bibinfo {year} {2022})},\ \Eprint
	{https://arxiv.org/abs/2205.06746} {arXiv:2205.06746 [hep-ph]} \BibitemShut
	{NoStop}%
	\bibitem [{\citenamefont {Li}(2024)}]{Li:2024fko}%
	\BibitemOpen
	\bibfield  {author} {\bibinfo {author} {\bibfnamefont {H.-n.}\ \bibnamefont
			{Li}},\ }\bibfield  {title} {\bibinfo {title} {{Dispersive Analysis of
				Excited Glueball States}},\ }\href
	{https://doi.org/10.1088/0256-307X/41/10/101101} {\bibfield  {journal}
		{\bibinfo  {journal} {Chin. Phys. Lett.}\ }\textbf {\bibinfo {volume} {41}},\
		\bibinfo {pages} {101101} (\bibinfo {year} {2024})},\ \Eprint
	{https://arxiv.org/abs/2408.06738} {arXiv:2408.06738 [hep-ph]} \BibitemShut
	{NoStop}%
	\bibitem [{\citenamefont {Zhao}\ \emph {et~al.}(2024)\citenamefont {Zhao},
		\citenamefont {Xing},\ and\ \citenamefont {Li}}]{Zhao:2024drr}%
	\BibitemOpen
	\bibfield  {author} {\bibinfo {author} {\bibfnamefont {Z.-X.}\ \bibnamefont
			{Zhao}}, \bibinfo {author} {\bibfnamefont {Y.-P.}\ \bibnamefont {Xing}},\
		and\ \bibinfo {author} {\bibfnamefont {R.-H.}\ \bibnamefont {Li}},\
	}\bibfield  {title} {\bibinfo {title} {{A progress in the inverse matrix
				method in QCD sum rules}},\ }\href
	{https://doi.org/10.1140/epjc/s10052-024-13452-8} {\bibfield  {journal}
		{\bibinfo  {journal} {Eur. Phys. J. C}\ }\textbf {\bibinfo {volume} {84}},\
		\bibinfo {pages} {1105} (\bibinfo {year} {2024})},\ \Eprint
	{https://arxiv.org/abs/2407.09819} {arXiv:2407.09819 [hep-ph]} \BibitemShut
	{NoStop}%
	\bibitem [{\citenamefont {Ou-Yang}\ \emph {et~al.}(2025)\citenamefont
		{Ou-Yang}, \citenamefont {Gubler}, \citenamefont {Oka}, \citenamefont
		{Wang},\ and\ \citenamefont {Wu}}]{Ou-Yang:2025efp}%
	\BibitemOpen
	\bibfield  {author} {\bibinfo {author} {\bibfnamefont {Z.-X.}\ \bibnamefont
			{Ou-Yang}}, \bibinfo {author} {\bibfnamefont {P.}~\bibnamefont {Gubler}},
		\bibinfo {author} {\bibfnamefont {M.}~\bibnamefont {Oka}}, \bibinfo {author}
		{\bibfnamefont {G.-J.}\ \bibnamefont {Wang}},\ and\ \bibinfo {author}
		{\bibfnamefont {J.-J.}\ \bibnamefont {Wu}},\ }\bibfield  {title} {\bibinfo
		{title} {{Resonance sum rules: An application to the square-well
				potential}},\ }\href {https://doi.org/10.1103/tcln-c924} {\bibfield
		{journal} {\bibinfo  {journal} {Phys. Rev. D}\ }\textbf {\bibinfo {volume}
			{112}},\ \bibinfo {pages} {114034} (\bibinfo {year} {2025})},\ \Eprint
	{https://arxiv.org/abs/2509.24336} {arXiv:2509.24336 [hep-ph]} \BibitemShut
	{NoStop}%
	\bibitem [{\citenamefont {Colangelo}\ and\ \citenamefont
		{Khodjamirian}(2000)}]{Colangelo:2000dp}%
	\BibitemOpen
	\bibfield  {author} {\bibinfo {author} {\bibfnamefont {P.}~\bibnamefont
			{Colangelo}}\ and\ \bibinfo {author} {\bibfnamefont {A.}~\bibnamefont
			{Khodjamirian}},\ }\bibfield  {title} {\bibinfo {title} {{QCD sum rules, a
				modern perspective}},\ }\href {https://doi.org/10.1142/9789812810458_0033} {\
		,\ \bibinfo {pages} {1495} (\bibinfo {year} {2000})},\ \Eprint
	{https://arxiv.org/abs/hep-ph/0010175} {arXiv:hep-ph/0010175} \BibitemShut
	{NoStop}%
	\bibitem [{\citenamefont {Kojo}\ \emph {et~al.}(2006)\citenamefont {Kojo},
		\citenamefont {Hayashigaki},\ and\ \citenamefont {Jido}}]{Kojo:2006bh}%
	\BibitemOpen
	\bibfield  {author} {\bibinfo {author} {\bibfnamefont {T.}~\bibnamefont
			{Kojo}}, \bibinfo {author} {\bibfnamefont {A.}~\bibnamefont {Hayashigaki}},\
		and\ \bibinfo {author} {\bibfnamefont {D.}~\bibnamefont {Jido}},\ }\bibfield
	{title} {\bibinfo {title} {{Pentaquark state in pole-dominated QCD sum
				rules}},\ }\href {https://doi.org/10.1103/PhysRevC.74.045206} {\bibfield
		{journal} {\bibinfo  {journal} {Phys. Rev. C}\ }\textbf {\bibinfo {volume}
			{74}},\ \bibinfo {pages} {045206} (\bibinfo {year} {2006})},\ \Eprint
	{https://arxiv.org/abs/hep-ph/0602004} {arXiv:hep-ph/0602004} \BibitemShut
	{NoStop}%
	\bibitem [{\citenamefont {Eden}\ \emph {et~al.}(1966)\citenamefont {Eden},
		\citenamefont {Landshoff}, \citenamefont {Olive},\ and\ \citenamefont
		{Polkinghorne}}]{Eden:1966dnq}%
	\BibitemOpen
	\bibfield  {author} {\bibinfo {author} {\bibfnamefont {R.~J.}\ \bibnamefont
			{Eden}}, \bibinfo {author} {\bibfnamefont {P.~V.}\ \bibnamefont {Landshoff}},
		\bibinfo {author} {\bibfnamefont {D.~I.}\ \bibnamefont {Olive}},\ and\
		\bibinfo {author} {\bibfnamefont {J.~C.}\ \bibnamefont {Polkinghorne}},\
	}\href@noop {} {\emph {\bibinfo {title} {{The analytic S-matrix}}}}\
	(\bibinfo  {publisher} {Cambridge Univ. Press},\ \bibinfo {address}
	{Cambridge},\ \bibinfo {year} {1966})\BibitemShut {NoStop}%
	\bibitem [{\citenamefont {Chew}(1962)}]{Chew:1962mpd}%
	\BibitemOpen
	\bibfield  {author} {\bibinfo {author} {\bibfnamefont {G.~F.}\ \bibnamefont
			{Chew}},\ }\bibfield  {title} {\bibinfo {title} {{S-Matrix Theory of Strong
				Interactions without Elementary Particles}},\ }\href
	{https://doi.org/10.1103/RevModPhys.34.394} {\bibfield  {journal} {\bibinfo
			{journal} {Rev. Mod. Phys.}\ }\textbf {\bibinfo {volume} {34}},\ \bibinfo
		{pages} {394} (\bibinfo {year} {1962})}\BibitemShut {NoStop}%
	\bibitem [{\citenamefont {Ben}\ \emph {et~al.}(2026)\citenamefont {Ben},
		\citenamefont {Yang},\ and\ \citenamefont {Zou}}]{Ben:2026afy}%
	\BibitemOpen
	\bibfield  {author} {\bibinfo {author} {\bibfnamefont {D.}~\bibnamefont
			{Ben}}, \bibinfo {author} {\bibfnamefont {L.-K.}\ \bibnamefont {Yang}},\ and\
		\bibinfo {author} {\bibfnamefont {B.-S.}\ \bibnamefont {Zou}},\ }\bibfield
	{title} {\bibinfo {title} {{On the Nature of $X(2370)$}},\ }\href@noop {} {\
		(\bibinfo {year} {2026})},\ \Eprint {https://arxiv.org/abs/2609.01342}
	{arXiv:2609.01342 [hep-ph]} \BibitemShut {NoStop}%
	\bibitem [{\citenamefont {Ablikim}\ \emph {et~al.}(2011)\citenamefont {Ablikim}
		\emph {et~al.}}]{BESIII:2010gmv}%
	\BibitemOpen
	\bibfield  {author} {\bibinfo {author} {\bibfnamefont {M.}~\bibnamefont
			{Ablikim}} \emph {et~al.} (\bibinfo {collaboration} {BESIII}),\ }\bibfield
	{title} {\bibinfo {title} {{Confirmation of the $X(1835)$ and observation of
				the resonances $X(2120)$ and $X(2370)$ in $J/\psi\to \gamma
				\pi^+\pi^-\eta^\prime$}},\ }\href
	{https://doi.org/10.1103/PhysRevLett.106.072002} {\bibfield  {journal}
		{\bibinfo  {journal} {Phys. Rev. Lett.}\ }\textbf {\bibinfo {volume} {106}},\
		\bibinfo {pages} {072002} (\bibinfo {year} {2011})},\ \Eprint
	{https://arxiv.org/abs/1012.3510} {arXiv:1012.3510 [hep-ex]} \BibitemShut
	{NoStop}%
	\bibitem [{\citenamefont {Ablikim}\ \emph {et~al.}(2020)\citenamefont {Ablikim}
		\emph {et~al.}}]{BESIII:2019wkp}%
	\BibitemOpen
	\bibfield  {author} {\bibinfo {author} {\bibfnamefont {M.}~\bibnamefont
			{Ablikim}} \emph {et~al.} (\bibinfo {collaboration} {BESIII}),\ }\bibfield
	{title} {\bibinfo {title} {{Observation of $X(2370)$ and search for X(2120)
				in $J/\psi \rightarrow \gamma K{\bar{K}} \eta '$}},\ }\href
	{https://doi.org/10.1140/epjc/s10052-020-8078-4} {\bibfield  {journal}
		{\bibinfo  {journal} {Eur. Phys. J. C}\ }\textbf {\bibinfo {volume} {80}},\
		\bibinfo {pages} {746} (\bibinfo {year} {2020})},\ \Eprint
	{https://arxiv.org/abs/1912.11253} {arXiv:1912.11253 [hep-ex]} \BibitemShut
	{NoStop}%
	\bibitem [{\citenamefont {Ablikim}\ \emph {et~al.}(2024)\citenamefont {Ablikim}
		\emph {et~al.}}]{BESIII:2023wfi}%
	\BibitemOpen
	\bibfield  {author} {\bibinfo {author} {\bibfnamefont {M.}~\bibnamefont
			{Ablikim}} \emph {et~al.} (\bibinfo {collaboration} {BESIII}),\ }\bibfield
	{title} {\bibinfo {title} {{Determination of Spin-Parity Quantum Numbers of
				X(2370) as 0-+ from
				J/{\ensuremath{\psi}}{\textrightarrow}{\ensuremath{\gamma}}KS0KS0{\ensuremath{\eta}}'}},\
	}\href {https://doi.org/10.1103/PhysRevLett.132.181901} {\bibfield  {journal}
		{\bibinfo  {journal} {Phys. Rev. Lett.}\ }\textbf {\bibinfo {volume} {132}},\
		\bibinfo {pages} {181901} (\bibinfo {year} {2024})},\ \Eprint
	{https://arxiv.org/abs/2312.05324} {arXiv:2312.05324 [hep-ex]} \BibitemShut
	{NoStop}%
	\bibitem [{\citenamefont {Lucha}\ \emph {et~al.}(2007)\citenamefont {Lucha},
		\citenamefont {Melikhov},\ and\ \citenamefont {Simula}}]{Lucha:2007pz}%
	\BibitemOpen
	\bibfield  {author} {\bibinfo {author} {\bibfnamefont {W.}~\bibnamefont
			{Lucha}}, \bibinfo {author} {\bibfnamefont {D.}~\bibnamefont {Melikhov}},\
		and\ \bibinfo {author} {\bibfnamefont {S.}~\bibnamefont {Simula}},\
	}\bibfield  {title} {\bibinfo {title} {{Systematic uncertainties of hadron
				parameters obtained with QCD sum rules}},\ }\href
	{https://doi.org/10.1103/PhysRevD.76.036002} {\bibfield  {journal} {\bibinfo
			{journal} {Phys. Rev. D}\ }\textbf {\bibinfo {volume} {76}},\ \bibinfo
		{pages} {036002} (\bibinfo {year} {2007})},\ \Eprint
	{https://arxiv.org/abs/0705.0470} {arXiv:0705.0470 [hep-ph]} \BibitemShut
	{NoStop}%
	\bibitem [{\citenamefont {Shifman}(2000)}]{Shifman:2000jv}%
	\BibitemOpen
	\bibfield  {author} {\bibinfo {author} {\bibfnamefont {M.~A.}\ \bibnamefont
			{Shifman}},\ }\bibfield  {title} {\bibinfo {title} {{Quark hadron duality}},\
	}in\ \href {https://doi.org/10.1142/9789812810458_0032} {\emph {\bibinfo
			{booktitle} {{8th International Symposium on Heavy Flavor Physics}}}},\
	Vol.~\bibinfo {volume} {3}\ (\bibinfo  {publisher} {World Scientific},\
	\bibinfo {address} {Singapore},\ \bibinfo {year} {2000})\ pp.\ \bibinfo
	{pages} {1447--1494},\ \Eprint {https://arxiv.org/abs/hep-ph/0009131}
	{arXiv:hep-ph/0009131} \BibitemShut {NoStop}%
	\bibitem [{\citenamefont {Shifman}(2010)}]{Shifman:2010zza}%
	\BibitemOpen
	\bibfield  {author} {\bibinfo {author} {\bibfnamefont {M.}~\bibnamefont
			{Shifman}},\ }\bibfield  {title} {\bibinfo {title} {{QCD sum rules: Bridging
				the gap between short and large distances}},\ }\href
	{https://doi.org/10.1016/j.nuclphysbps.2010.10.075} {\bibfield  {journal}
		{\bibinfo  {journal} {Nucl. Phys. B Proc. Suppl.}\ }\textbf {\bibinfo
			{volume} {207-208}},\ \bibinfo {pages} {298} (\bibinfo {year} {2010})},\
	\Eprint {https://arxiv.org/abs/1101.1122} {arXiv:1101.1122 [hep-ph]}
	\BibitemShut {NoStop}%
	\bibitem [{\citenamefont {Albuquerque}(2026)}]{Albuquerque:2026gll}%
	\BibitemOpen
	\bibfield  {author} {\bibinfo {author} {\bibfnamefont {R.~M.}\ \bibnamefont
			{Albuquerque}},\ }\bibfield  {title} {\bibinfo {title} {{Beyond Borel
				Windows: A Systematic Optimization Framework for QCD Sum Rules}},\ }in\
	\href@noop {} {\emph {\bibinfo {booktitle} {{29th High-Energy Physics
					International Conference in Quantum Chromodynamics}: {QCD26}}}}\ (\bibinfo
	{year} {2026})\ \Eprint {https://arxiv.org/abs/2608.05322} {arXiv:2608.05322
		[hep-ph]} \BibitemShut {NoStop}%
	\bibitem [{\citenamefont {Wunderlich}\ \emph {et~al.}(2026)\citenamefont
		{Wunderlich}, \citenamefont {Bulgakov}, \citenamefont {Joo}, \citenamefont
		{Lee},\ and\ \citenamefont {Mokeev}}]{Wunderlich:2026krs}%
	\BibitemOpen
	\bibfield  {author} {\bibinfo {author} {\bibfnamefont {Y.}~\bibnamefont
			{Wunderlich}}, \bibinfo {author} {\bibfnamefont {A.}~\bibnamefont
			{Bulgakov}}, \bibinfo {author} {\bibfnamefont {K.}~\bibnamefont {Joo}},
		\bibinfo {author} {\bibfnamefont {T.~S.~H.}\ \bibnamefont {Lee}},\ and\
		\bibinfo {author} {\bibfnamefont {V.~I.}\ \bibnamefont {Mokeev}},\ }\bibfield
	{title} {\bibinfo {title} {{Quark--hadron duality in inclusive
				electron--proton scattering at high $Q^{2}$: structure functions and
				truncated moments from CLAS12}},\ }\href@noop {} {\  (\bibinfo {year}
		{2026})},\ \Eprint {https://arxiv.org/abs/2606.29690} {arXiv:2606.29690
		[hep-ph]} \BibitemShut {NoStop}%
	\bibitem [{\citenamefont {Bloom}\ and\ \citenamefont
		{Gilman}(1970)}]{Bloom:1970xb}%
	\BibitemOpen
	\bibfield  {author} {\bibinfo {author} {\bibfnamefont {E.~D.}\ \bibnamefont
			{Bloom}}\ and\ \bibinfo {author} {\bibfnamefont {F.~J.}\ \bibnamefont
			{Gilman}},\ }\bibfield  {title} {\bibinfo {title} {{Scaling, Duality, and the
				Behavior of Resonances in Inelastic electron-Proton Scattering}},\ }\href
	{https://doi.org/10.1103/PhysRevLett.25.1140} {\bibfield  {journal} {\bibinfo
			{journal} {Phys. Rev. Lett.}\ }\textbf {\bibinfo {volume} {25}},\ \bibinfo
		{pages} {1140} (\bibinfo {year} {1970})}\BibitemShut {NoStop}%
	\bibitem [{\citenamefont {Bloom}\ and\ \citenamefont
		{Gilman}(1971)}]{Bloom:1971ye}%
	\BibitemOpen
	\bibfield  {author} {\bibinfo {author} {\bibfnamefont {E.~D.}\ \bibnamefont
			{Bloom}}\ and\ \bibinfo {author} {\bibfnamefont {F.~J.}\ \bibnamefont
			{Gilman}},\ }\bibfield  {title} {\bibinfo {title} {{Scaling and the Behavior
				of Nucleon Resonances in Inelastic electron-Nucleon Scattering}},\ }\href
	{https://doi.org/10.1103/PhysRevD.4.2901} {\bibfield  {journal} {\bibinfo
			{journal} {Phys. Rev. D}\ }\textbf {\bibinfo {volume} {4}},\ \bibinfo {pages}
		{2901} (\bibinfo {year} {1971})}\BibitemShut {NoStop}%
	\bibitem [{\citenamefont {Melnitchouk}\ \emph {et~al.}(2005)\citenamefont
		{Melnitchouk}, \citenamefont {Ent},\ and\ \citenamefont
		{Keppel}}]{Melnitchouk:2005zr}%
	\BibitemOpen
	\bibfield  {author} {\bibinfo {author} {\bibfnamefont {W.}~\bibnamefont
			{Melnitchouk}}, \bibinfo {author} {\bibfnamefont {R.}~\bibnamefont {Ent}},\
		and\ \bibinfo {author} {\bibfnamefont {C.}~\bibnamefont {Keppel}},\
	}\bibfield  {title} {\bibinfo {title} {{Quark-hadron duality in electron
				scattering}},\ }\href {https://doi.org/10.1016/j.physrep.2004.10.004}
	{\bibfield  {journal} {\bibinfo  {journal} {Phys. Rept.}\ }\textbf {\bibinfo
			{volume} {406}},\ \bibinfo {pages} {127} (\bibinfo {year} {2005})},\ \Eprint
	{https://arxiv.org/abs/hep-ph/0501217} {arXiv:hep-ph/0501217} \BibitemShut
	{NoStop}%
	\bibitem [{\citenamefont {Malace}\ \emph {et~al.}(2009)\citenamefont {Malace}
		\emph {et~al.}}]{JeffersonLabE00-115:2009jll}%
	\BibitemOpen
	\bibfield  {author} {\bibinfo {author} {\bibfnamefont {S.~P.}\ \bibnamefont
			{Malace}} \emph {et~al.} (\bibinfo {collaboration} {Jefferson Lab E00-115}),\
	}\bibfield  {title} {\bibinfo {title} {{Applications of quark-hadron duality
				in F(2) structure function}},\ }\href
	{https://doi.org/10.1103/PhysRevC.80.035207} {\bibfield  {journal} {\bibinfo
			{journal} {Phys. Rev. C}\ }\textbf {\bibinfo {volume} {80}},\ \bibinfo
		{pages} {035207} (\bibinfo {year} {2009})},\ \Eprint
	{https://arxiv.org/abs/0905.2374} {arXiv:0905.2374 [nucl-ex]} \BibitemShut
	{NoStop}%
	\bibitem [{\citenamefont {Zhang}\ \emph {et~al.}(2025)\citenamefont {Zhang},
		\citenamefont {Chen}, \citenamefont {Ou-Yang}, \citenamefont {Yu},
		\citenamefont {Chen},\ and\ \citenamefont {Wu}}]{Zhang:2025fuz}%
	\BibitemOpen
	\bibfield  {author} {\bibinfo {author} {\bibfnamefont {J.-P.}\ \bibnamefont
			{Zhang}}, \bibinfo {author} {\bibfnamefont {X.-L.}\ \bibnamefont {Chen}},
		\bibinfo {author} {\bibfnamefont {Z.-X.}\ \bibnamefont {Ou-Yang}}, \bibinfo
		{author} {\bibfnamefont {X.}~\bibnamefont {Yu}}, \bibinfo {author}
		{\bibfnamefont {W.}~\bibnamefont {Chen}},\ and\ \bibinfo {author}
		{\bibfnamefont {J.-J.}\ \bibnamefont {Wu}},\ }\bibfield  {title} {\bibinfo
		{title} {{Unraveling K(1690) as a pseudoscalar
				udd{\textasciimacron}s{\textasciimacron} tetraquark state}},\ }\href
	{https://doi.org/10.1103/fv87-6bcr} {\bibfield  {journal} {\bibinfo
			{journal} {Phys. Rev. D}\ }\textbf {\bibinfo {volume} {112}},\ \bibinfo
		{pages} {094047} (\bibinfo {year} {2025})},\ \Eprint
	{https://arxiv.org/abs/2507.05726} {arXiv:2507.05726 [hep-ph]} \BibitemShut
	{NoStop}%
	\bibitem [{\citenamefont {Navas}\ \emph {et~al.}(2024)\citenamefont {Navas}
		\emph {et~al.}}]{ParticleDataGroup:2024cfk}%
	\BibitemOpen
	\bibfield  {author} {\bibinfo {author} {\bibfnamefont {S.}~\bibnamefont
			{Navas}} \emph {et~al.} (\bibinfo {collaboration} {Particle Data Group}),\
	}\bibfield  {title} {\bibinfo {title} {{Review of particle physics}},\ }\href
	{https://doi.org/10.1103/PhysRevD.110.030001} {\bibfield  {journal} {\bibinfo
			{journal} {Phys. Rev. D}\ }\textbf {\bibinfo {volume} {110}},\ \bibinfo
		{pages} {030001} (\bibinfo {year} {2024})}\BibitemShut {NoStop}%
	\bibitem [{\citenamefont {Narison}(2018)}]{Narison:2018dcr}%
	\BibitemOpen
	\bibfield  {author} {\bibinfo {author} {\bibfnamefont {S.}~\bibnamefont
			{Narison}},\ }\bibfield  {title} {\bibinfo {title} {{QCD parameter
				correlations from heavy quarkonia}},\ }\href
	{https://doi.org/10.1142/S0217751X18500458} {\bibfield  {journal} {\bibinfo
			{journal} {Int. J. Mod. Phys. A}\ }\textbf {\bibinfo {volume} {33}},\
		\bibinfo {pages} {1850045} (\bibinfo {year} {2018})},\ \bibinfo {note}
	{[Addendum: Int.J.Mod.Phys.A 33, 1892004 (2018)]},\ \Eprint
	{https://arxiv.org/abs/1801.00592} {arXiv:1801.00592 [hep-ph]} \BibitemShut
	{NoStop}%
	\bibitem [{\citenamefont {Narison}(2025)}]{Narison:2025cys}%
	\BibitemOpen
	\bibfield  {author} {\bibinfo {author} {\bibfnamefont {S.}~\bibnamefont
			{Narison}},\ }\bibfield  {title} {\bibinfo {title} {{QCD condensates and
				{\ensuremath{\alpha}}s from {\ensuremath{\tau}}-decay: Summary}},\ }\href
	{https://doi.org/10.1016/j.jspc.2025.100038} {\bibfield  {journal} {\bibinfo
			{journal} {J. Subatomic Part. Cosmol.}\ }\textbf {\bibinfo {volume} {3}},\
		\bibinfo {pages} {100038} (\bibinfo {year} {2025})},\ \Eprint
	{https://arxiv.org/abs/2501.08369} {arXiv:2501.08369 [hep-ph]} \BibitemShut
	{NoStop}%
	\bibitem [{\citenamefont {Jin}\ \emph {et~al.}(2003)\citenamefont {Jin},
		\citenamefont {Korner},\ and\ \citenamefont {Steele}}]{Jin:2002rw}%
	\BibitemOpen
	\bibfield  {author} {\bibinfo {author} {\bibfnamefont {H.~Y.}\ \bibnamefont
			{Jin}}, \bibinfo {author} {\bibfnamefont {J.~G.}\ \bibnamefont {Korner}},\
		and\ \bibinfo {author} {\bibfnamefont {T.~G.}\ \bibnamefont {Steele}},\
	}\bibfield  {title} {\bibinfo {title} {{Improved determination of the mass of
				the 1-+ light hybrid meson from QCD sum rules}},\ }\href
	{https://doi.org/10.1103/PhysRevD.67.014025} {\bibfield  {journal} {\bibinfo
			{journal} {Phys. Rev. D}\ }\textbf {\bibinfo {volume} {67}},\ \bibinfo
		{pages} {014025} (\bibinfo {year} {2003})},\ \Eprint
	{https://arxiv.org/abs/hep-ph/0211304} {arXiv:hep-ph/0211304} \BibitemShut
	{NoStop}%
	\bibitem [{\citenamefont {Huang}\ \emph {et~al.}(2015)\citenamefont {Huang},
		\citenamefont {Jin},\ and\ \citenamefont {Zhang}}]{Huang:2014hya}%
	\BibitemOpen
	\bibfield  {author} {\bibinfo {author} {\bibfnamefont {Z.-R.}\ \bibnamefont
			{Huang}}, \bibinfo {author} {\bibfnamefont {H.-Y.}\ \bibnamefont {Jin}},\
		and\ \bibinfo {author} {\bibfnamefont {Z.-F.}\ \bibnamefont {Zhang}},\
	}\bibfield  {title} {\bibinfo {title} {{New predictions on the mass of the
				$1^{-+}$ light hybrid meson from QCD sum rules}},\ }\href
	{https://doi.org/10.1007/JHEP04(2015)004} {\bibfield  {journal} {\bibinfo
			{journal} {JHEP}\ }\textbf {\bibinfo {volume} {04}},\ \bibinfo {pages}
		{004}},\ \Eprint {https://arxiv.org/abs/1411.2224} {arXiv:1411.2224 [hep-ph]}
	\BibitemShut {NoStop}%
	\bibitem [{\citenamefont {Di~Giacomo}\ \emph {et~al.}(1982)\citenamefont
		{Di~Giacomo}, \citenamefont {Fabricius},\ and\ \citenamefont
		{Paffuti}}]{DiGiacomo:1982gn}%
	\BibitemOpen
	\bibfield  {author} {\bibinfo {author} {\bibfnamefont {A.}~\bibnamefont
			{Di~Giacomo}}, \bibinfo {author} {\bibfnamefont {K.}~\bibnamefont
			{Fabricius}},\ and\ \bibinfo {author} {\bibfnamefont {G.}~\bibnamefont
			{Paffuti}},\ }\bibfield  {title} {\bibinfo {title} {{Trilinear Gluon
				Condensation Parameter From Lattice {QCD}}},\ }\href
	{https://doi.org/10.1016/0370-2693(82)90615-3} {\bibfield  {journal}
		{\bibinfo  {journal} {Phys. Lett. B}\ }\textbf {\bibinfo {volume} {118}},\
		\bibinfo {pages} {129} (\bibinfo {year} {1982})}\BibitemShut {NoStop}%
	\bibitem [{\citenamefont {Ioffe}\ \emph {et~al.}(2010)\citenamefont {Ioffe},
		\citenamefont {Fadin},\ and\ \citenamefont {Lipatov}}]{Ioffe:2010zz}%
	\BibitemOpen
	\bibfield  {author} {\bibinfo {author} {\bibfnamefont {B.~L.}\ \bibnamefont
			{Ioffe}}, \bibinfo {author} {\bibfnamefont {V.~S.}\ \bibnamefont {Fadin}},\
		and\ \bibinfo {author} {\bibfnamefont {L.~N.}\ \bibnamefont {Lipatov}},\
	}\href {https://doi.org/10.1017/CBO9780511711817} {\emph {\bibinfo {title}
			{{Quantum chromodynamics: Perturbative and nonperturbative aspects}}}}\
	(\bibinfo  {publisher} {Cambridge Univ. Press},\ \bibinfo {year}
	{2010})\BibitemShut {NoStop}%
	\bibitem [{\citenamefont {Narison}(2012{\natexlab{a}})}]{Narison:2011rn}%
	\BibitemOpen
	\bibfield  {author} {\bibinfo {author} {\bibfnamefont {S.}~\bibnamefont
			{Narison}},\ }\bibfield  {title} {\bibinfo {title} {{Gluon Condensates and
				$\bar{m}_b(\bar{m}_b)$ from QCD-Exponential Moments at Higher Orders}},\
	}\href {https://doi.org/10.1016/j.physletb.2011.12.047} {\bibfield  {journal}
		{\bibinfo  {journal} {Phys. Lett. B}\ }\textbf {\bibinfo {volume} {707}},\
		\bibinfo {pages} {259} (\bibinfo {year} {2012}{\natexlab{a}})},\ \Eprint
	{https://arxiv.org/abs/1105.5070} {arXiv:1105.5070 [hep-ph]} \BibitemShut
	{NoStop}%
	\bibitem [{\citenamefont {Narison}(2012{\natexlab{b}})}]{Narison:2011xe}%
	\BibitemOpen
	\bibfield  {author} {\bibinfo {author} {\bibfnamefont {S.}~\bibnamefont
			{Narison}},\ }\bibfield  {title} {\bibinfo {title} {{Gluon Condensates and
				precise $\overline{m}_{c,b}$ from QCD-Moments and their ratios to Order
				$\alpha_s^3$ and {\ensuremath{<}} G$^4$ {\ensuremath{>}}}},\ }\href
	{https://doi.org/10.1016/j.physletb.2011.11.058} {\bibfield  {journal}
		{\bibinfo  {journal} {Phys. Lett. B}\ }\textbf {\bibinfo {volume} {706}},\
		\bibinfo {pages} {412} (\bibinfo {year} {2012}{\natexlab{b}})},\ \Eprint
	{https://arxiv.org/abs/1105.2922} {arXiv:1105.2922 [hep-ph]} \BibitemShut
	{NoStop}%
	\bibitem [{\citenamefont {Ablikim}\ \emph {et~al.}(2026)\citenamefont {Ablikim}
		\emph {et~al.}}]{BESIII:2026emk}%
	\BibitemOpen
	\bibfield  {author} {\bibinfo {author} {\bibfnamefont {M.}~\bibnamefont
			{Ablikim}} \emph {et~al.} (\bibinfo {collaboration} {BESIII}),\ }\bibfield
	{title} {\bibinfo {title} {{Measurements of the branching fractions of
				\(\chi_{cJ} \to \Sigma^0 \bar{\Sigma}^0 \eta\)}},\ }\href@noop {} {\
		(\bibinfo {year} {2026})},\ \Eprint {https://arxiv.org/abs/2610.15332}
	{arXiv:2610.15332 [hep-ex]} \BibitemShut {NoStop}%
	\bibitem [{\citenamefont {Chung}\ \emph {et~al.}(1981)\citenamefont {Chung},
		\citenamefont {Dosch}, \citenamefont {Kremer},\ and\ \citenamefont
		{Schall}}]{Chung:1981wm}%
	\BibitemOpen
	\bibfield  {author} {\bibinfo {author} {\bibfnamefont {Y.}~\bibnamefont
			{Chung}}, \bibinfo {author} {\bibfnamefont {H.~G.}\ \bibnamefont {Dosch}},
		\bibinfo {author} {\bibfnamefont {M.}~\bibnamefont {Kremer}},\ and\ \bibinfo
		{author} {\bibfnamefont {D.}~\bibnamefont {Schall}},\ }\bibfield  {title}
	{\bibinfo {title} {{{QCD} Sum Rules for 'Baryonic Currents'}},\ }\href
	{https://doi.org/10.1016/0370-2693(81)91057-1} {\bibfield  {journal}
		{\bibinfo  {journal} {Phys. Lett. B}\ }\textbf {\bibinfo {volume} {102}},\
		\bibinfo {pages} {175} (\bibinfo {year} {1981})}\BibitemShut {NoStop}%
	\bibitem [{\citenamefont {Wan}\ \emph {et~al.}(2022)\citenamefont {Wan},
		\citenamefont {Zhang},\ and\ \citenamefont {Qiao}}]{Wan:2021vny}%
	\BibitemOpen
	\bibfield  {author} {\bibinfo {author} {\bibfnamefont {B.-D.}\ \bibnamefont
			{Wan}}, \bibinfo {author} {\bibfnamefont {S.-Q.}\ \bibnamefont {Zhang}},\
		and\ \bibinfo {author} {\bibfnamefont {C.-F.}\ \bibnamefont {Qiao}},\
	}\bibfield  {title} {\bibinfo {title} {{Light baryonium spectrum}},\ }\href
	{https://doi.org/10.1103/PhysRevD.105.014016} {\bibfield  {journal} {\bibinfo
			{journal} {Phys. Rev. D}\ }\textbf {\bibinfo {volume} {105}},\ \bibinfo
		{pages} {014016} (\bibinfo {year} {2022})},\ \Eprint
	{https://arxiv.org/abs/2109.07130} {arXiv:2109.07130 [hep-ph]} \BibitemShut
	{NoStop}%
	\bibitem [{\citenamefont {Xu}\ \emph {et~al.}(2025)\citenamefont {Xu},
		\citenamefont {Chen}, \citenamefont {Zhang}, \citenamefont {Li},\ and\
		\citenamefont {Chen}}]{Xu:2025oqn}%
	\BibitemOpen
	\bibfield  {author} {\bibinfo {author} {\bibfnamefont {G.-F.}\ \bibnamefont
			{Xu}}, \bibinfo {author} {\bibfnamefont {X.-L.}\ \bibnamefont {Chen}},
		\bibinfo {author} {\bibfnamefont {J.-P.}\ \bibnamefont {Zhang}}, \bibinfo
		{author} {\bibfnamefont {N.}~\bibnamefont {Li}},\ and\ \bibinfo {author}
		{\bibfnamefont {W.}~\bibnamefont {Chen}},\ }\bibfield  {title} {\bibinfo
		{title} {{Possible bound states in the triple-$\eta_c$ and triple-$J/\psi$
				systems}},\ }\href {https://doi.org/10.1088/0256-307X/42/7/070201} {\bibfield
		{journal} {\bibinfo  {journal} {Chin. Phys. Lett.}\ }\textbf {\bibinfo
			{volume} {42}},\ \bibinfo {pages} {070201} (\bibinfo {year} {2025})},\
	\Eprint {https://arxiv.org/abs/2504.08665} {arXiv:2504.08665 [hep-ph]}
	\BibitemShut {NoStop}%
	\bibitem [{\citenamefont {Li}\ \emph {et~al.}(2025)\citenamefont {Li},
		\citenamefont {Lai},\ and\ \citenamefont {Jin}}]{Li:2025dkw}%
	\BibitemOpen
	\bibfield  {author} {\bibinfo {author} {\bibfnamefont {S.-H.}\ \bibnamefont
			{Li}}, \bibinfo {author} {\bibfnamefont {W.-Y.}\ \bibnamefont {Lai}},\ and\
		\bibinfo {author} {\bibfnamefont {H.-Y.}\ \bibnamefont {Jin}},\ }\bibfield
	{title} {\bibinfo {title} {{Probing Excited $q\bar{q}$ Mesons via QCD Sum
				Rules}},\ }\href@noop {} {\  (\bibinfo {year} {2025})},\ \Eprint
	{https://arxiv.org/abs/2512.16637} {arXiv:2512.16637 [hep-ph]} \BibitemShut
	{NoStop}%
	\bibitem [{\citenamefont {Li}\ \emph {et~al.}(2026)\citenamefont {Li},
		\citenamefont {Huang}, \citenamefont {Chen},\ and\ \citenamefont
		{Jin}}]{Li:2025hsp}%
	\BibitemOpen
	\bibfield  {author} {\bibinfo {author} {\bibfnamefont {S.-H.}\ \bibnamefont
			{Li}}, \bibinfo {author} {\bibfnamefont {Z.-R.}\ \bibnamefont {Huang}},
		\bibinfo {author} {\bibfnamefont {W.}~\bibnamefont {Chen}},\ and\ \bibinfo
		{author} {\bibfnamefont {H.-Y.}\ \bibnamefont {Jin}},\ }\bibfield  {title}
	{\bibinfo {title} {{Revising the mass of light hybrid mesons: NLO QCD sum
				rules point to {\ensuremath{\phi}}(2170) as a prime candidate}},\ }\href
	{https://doi.org/10.1007/JHEP03(2026)087} {\bibfield  {journal} {\bibinfo
			{journal} {JHEP}\ }\textbf {\bibinfo {volume} {03}},\ \bibinfo {pages}
		{087}},\ \Eprint {https://arxiv.org/abs/2506.22412} {arXiv:2506.22412
		[hep-ph]} \BibitemShut {NoStop}%
	\bibitem [{\citenamefont {Chen}\ \emph {et~al.}(2024)\citenamefont {Chen},
		\citenamefont {Chen}, \citenamefont {Yang},\ and\ \citenamefont
		{Chen}}]{Chen:2024bpz}%
	\BibitemOpen
	\bibfield  {author} {\bibinfo {author} {\bibfnamefont {Z.-Z.}\ \bibnamefont
			{Chen}}, \bibinfo {author} {\bibfnamefont {X.-L.}\ \bibnamefont {Chen}},
		\bibinfo {author} {\bibfnamefont {P.-F.}\ \bibnamefont {Yang}},\ and\
		\bibinfo {author} {\bibfnamefont {W.}~\bibnamefont {Chen}},\ }\bibfield
	{title} {\bibinfo {title} {{P-wave fully charm and fully bottom tetraquark
				states}},\ }\href {https://doi.org/10.1103/PhysRevD.109.094011} {\bibfield
		{journal} {\bibinfo  {journal} {Phys. Rev. D}\ }\textbf {\bibinfo {volume}
			{109}},\ \bibinfo {pages} {094011} (\bibinfo {year} {2024})},\ \Eprint
	{https://arxiv.org/abs/2402.03117} {arXiv:2402.03117 [hep-ph]} \BibitemShut
	{NoStop}%
\end{thebibliography}
\end{document}